\documentclass[preprint,showpacs,showkeys,aps,prd,amsfonts,eqsecnum,nofootinbib,
floatfix]
{revtex4-1}
\usepackage{graphicx,epsfig}
\usepackage[dvips]{color}
\def\gsim{\lower0.5ex\hbox{$\:\buildrel >\over\sim\:$}}
\def\lsim{\lower0.5ex\hbox{$\:\buildrel <\over\sim\:$}}
\begin{document}
\preprint{CUMQ/HEP 150}

\title{\Large Production and Decays of $W_R$ bosons at the LHC}
\author{Mariana Frank$^{a}$}
\author{Alper Hayreter$^{a}$} 
\author{Ismail Turan$^{b}$ }

\affiliation{$^{a}$Department of Physics, Concordia University, 7141 Sherbrooke St. 
West, Montreal, Quebec, CANADA H4B 1R6,}
 \affiliation{$^{b}$Ottawa-Carleton 
Institute of Physics,
Carleton University,
1125 Colonel By Drive
Ottawa, Ontario, Canada, K1S 5B6.}

\date{\today}

\begin{abstract}
With the advent of the LHC, it is important to devise clear tests for Physics
 Beyond the Standard Model. Such physics could manifest itself in the form of
new charged bosons, whose presence is most naturally occurring in left-right
symmetric models (LRSM). We analyze  the single $W_R$ boson production in an
asymmetric left-right model, where the left and right quark mixing matrices are
not constrained to be equal. We investigate the cross sections as well as
branching ratios of $W_R$ bosons at the LHC, including constraints from low
energy phenomenology. We then look for most likely signals in $pp \to W_R\ t \to
t ~(dijet)$ production. Including the background, we find that LHC could show
significant signals for the new charged bosons. We compare our results
throughout with the manifest left-right symmetric model and comment on
similarities and differences.

\end{abstract}
\maketitle
 
 \section{Introduction}
 \label{sec:intro}
 
While the Standard Model (SM) has provided a  compelling picture of low energy
 interactions, it has been plagued by theoretical inconsistencies. More
recently, experimental deviations from the predictions of the model (such as
signals of neutrino masses and mixing \cite{Fukuda:1998mi}) have given further
justification to building model of Physics Beyond the Standard Model.
Additionally, the experimental outlook on testing these scenarios looks very
promising. LHC data  is expected to provide ample material for analysis. When
the data becomes available, it would be difficult to disentangle expectations
for different models. The task of theorists is to provide viable scenarios for
Physics Beyond the Standard Model {\it and} to predict the signals which
distinguish them from the SM and from each other.

A large variety of models is available, all of which attempt to resolve some
 theoretical inconsistency of the SM. Of these, a particularly simple model is
the Left Right Symmetric Model (LRSM) \cite{Pati:1974yy}. Originally introduced
to resolve the parity and neutrino mass problems, it remains one of the simplest
extensions of the SM, and it is a natural scenario for the seesaw mechanism
\cite{Mohapatra:1979ia}. The Higgs sector of the LRSM and its signals at
accelerators have been thoroughly analyzed by theorists \cite{Gunion:1989in},
and experimentalists have been particularly keen to search for doubly charged
Higgs bosons, predicted in most versions of the model \cite{Abazov:2004au} .
Less attention has been paid to the vector boson sector. The  LRSM extends the
gauge group of the SM to $SU(2)_L \times SU(2)_R \times U(1)_{B-L}$, and thus
predicts the existence of two extra gauge bosons: a neutral $Z_R$ and a charged
$W_R$. While an extra neutral gauge boson $Z^{\prime}$  is predicted by several
extensions of the SM, all containing an extra gauged $U(1)$ symmetry
group, a charged gauge boson would be a more likely indication of left right
symmetry\footnote{While the existence of $W_R$ is present in several gauge
unification scenarios,  models with extra $W_L$ bosons also exist.}. 

Several other models predict the existence of extra $W^\prime$ bosons, such as
 extra dimensional models (both Randall Sundrum  \cite{RS} and Universal Extra
Dimensions model \cite{UED}), Little Higgs models \cite{Little Higgs} and
Composite Higgs models \cite{Higgless}.  The  $W^\prime$ predicted in these
models  have features which distinguish them from LRSM, which we will discuss
after our analysis.

Production of extra charged vector bosons at colliders has received less
interest
 than that of $Z^\prime$, although one study exists for the Tevatron \cite{Rizzo:1993fe}. However, recently, papers have appeared which analyze
chiral couplings of a $W^\prime$ at LHC and indicate how to disentangle left or
right handed bosons \cite{Gopalakrishna:2010xm}. In the present work, we follow
a different procedure. Assuming the extra charged vector boson to come from a
version of LRSM (and thus be right-handed), we analyze the production mechanism,
decay rates and possible signals at the LHC. 

$W_R$ bosons are predicted to be heavy, of $\cal{O}$(TeV) and thus the signal
 expected to be much below the $W_L$ production signal. But this is only the
case if the quark mixing matrix in the right-handed sector ($V_{CKM}^R$) is
either identical, or equal up to a diagonal matrix, to the one in the
left-handed sector (the usual Cabibbo-Kobayashi-Maskawa $V_{CKM}^L$)--the so
called {\it manifest} and {\it pseudo-manifest} LRSM, respectively. We describe the distinctive features of these models in the next section.  This does not
have to be the case, as was discussed at length by Langacker and Sankar
\cite{Langacker:1989xa}, who allow right handed mixing matrices $V_{CKM}^R$ with
large off-diagonal elements.
They perform a thorough investigation of the constraints on the mass of $W_R$
 and its mixing with $W_L$ under these circumstances, and find out that the $W_R$
mass can be a lot lighter, $M_{W_R} > 300$ GeV \cite{Amsler:2008zz}. In what follows, we refer to this model and its variants as the asymmetric left-right model.

In the age of LHC, there is another immediate advantage of the asymmetric left-right model: such
 a $W_R$ boson can be produced at rates larger by orders of magnitude than for
models in which $V_{CKM}^R= V_{CKM}^{L \star}K$, where $K$ is a diagonal phase
matrix.  One could see this by looking at the signal $pp \to W_{L,R} t$. This
single-top production cross section is known to be important in identifying and
distinguishing between different new physics models, as these  can have
different effects ($s$-channel or $t$-channel) on the production process
\cite{Tait:2000sh}. The partonic cross section at LHC is dominated by $q g$,
with $q=d,s$. However for $W_L$ production one must rely on the process $g b \to
b \to t W $, and thus be disadvantaged by the small amount of $b$ quarks in the
proton; or rely on $g d(s) \to d(s) \to t W$, which is suppressed by the
$V^L_{ts}$ or $V^L_{td}$ element of the $V_{CKM}^L$. However, if the
off-diagonal $V^R_{ts}$ or $V^R_{td}$ elements of the $V_{CKM}^R$ are large, one
could produce $W_R$ copiously. Additionally if there are  less stringent
restrictions on the $W_R$ mass, one can envisage that $W_R$ production could be
observable, and if so, a clear distinguishing signal for LRSM.  At the Tevatron,
the production cross section is dominated at the partonic level by $q {\bar q}$,
with $q=u,\,d,\, s,\, c$.  Even for a light $W_R$ boson, we would not expect any
enhancements due to non-diagonal entries in the $V_{CKM}^R$, and the same is
true for linear colliders. 

The sensitivity of the Tevatron to $W_R$ searches has been thoroughly discussed in  \cite{Rizzo:1993fe}. Mass limits from the existing data depend on the ratio of the coupling constants for $SU(2)_R$ and $SU(2)_L$, $g_R/g_L$, on the nature and mass of the right-handed neutrinos $\nu_R$, on the leptonic branching ratio for $W_R$, on the form of the right-handed CKM matrix $V^R_{CKM}$. The most stringent experimental bounds from Tevatron searches are $M_{W_R} \ge 1$ TeV, under very specific assumptions (looking for $W_R$ decays into an electron and a neutrino, for Standard Model-like couplings to fermions) \cite{Tevatron}.  As their assumptions would not  apply for our model, we investigate the possible signals and mass bounds at the Tevatron in $dijet$ production before proceeding with the LHC signal analysis. 

The LHC thus presents a unique opportunity to observe
such a $W_R$ boson. We propose to investigate this possibility in the present
paper. 
In a previous paper \cite{Frank:2010qv}, we have laid the foundation of
 flavor-changing studies in left-right models by analyzing the most general
restrictions on the parameter space of the model ($M_{W_R}, M_{H^\pm},
V_{CKM}^R$, and $g_R/g_L$) coming from $ b \to s \gamma$, $B^0_d- {\bar B}^0_d$
and $B^0_s- {\bar B}^0_s$  mixings. For consistency, we include here these
parameter space restrictions, as well as those coming from the Kaon physics.

We will proceed as follows. In Section \ref{lrm} we will review the existing
 models and define the free parameters.  We then briefly summarize the
constraints on the $W_R$ mass and mixing coming from $K$ and $B$ meson
phenomenology in Section \ref{constraints}. We then investigate the production
and decays of $W_R$ bosons at the LHC under different $V_{CKM}^R$
parametrizations, and indicate which type can produce the most promising signals
in Section \ref{prod-dec}. There we analyze the background and give an idea of
the signals expected at LHC. We conclude and summarize our findings in Section
\ref{conclude}.

\section{Left-Right Symmetric Models}
\label{lrm}

The idea of the original model was to construct a model based on
 $SU(2)_L \times SU(2)_R \times U(1)$ whose Lagrangian was invariant under a
discrete left-right symmetry \cite{Pati:1974yy}. This implied that the gauge
couplings $g_L$ and $g_R$ were equal and the Yukawa couplings were also
restricted. Parity violation occurred through spontaneous symmetry breaking and
provided different masses for $W_L$ and $W_R$ bosons\footnote{These models run
into difficulty because they predict  $\sin^2 \theta_W$ too large and have
problems with the baryon asymmetry and cosmological domain walls
\cite{Langacker:1989xa}.}.

Then the idea of discrete left-right (LR) symmetry was explored,
 where LR symmetry was assumed broken at a higher scale than the $SU(2)_L \times
SU(2)_R \times U(1)$ breaking scale, which allows $g_L \ne g_R$.   This led to
{\it manifest LR symmetric models} \cite{Senjanovic:1978ev}, where the CP
violation is generated by complex Yukawa couplings, while the vevs of the Higgs
fields remain real. This implies the same mixing for right and left-handed
quarks, $V_{CKM}^R= V_{CKM}^L$, where $V_{CKM}^L$ is the usual
Cabibbo-Kobayashi-Maskawa matrix.

In {\it pseudo-manifest LR symmetry} both CP and P symmetries are spontaneously
 broken \cite{Harari:1983gq}, such that the Yukawa couplings are real. In that
case the left and right handed quark mixings are related through $V_{CKM}^R=
V_{CKM}^{L \star}K$, with $K$ a diagonal phase matrix. Since in this model CP is
spontaneously broken, this scenario shares the problems of the model invariant
under discrete LR symmetry.

As problems seem to be consequences of requiring $V_{CKM}^L= V_{CKM}^R$,
 Langacker and Sankar have abandoned it in favor of a more general LR model 
\cite{Langacker:1989xa} (which we call the asymmetric model, because mixings in
the left and right handed quark sectors are not required to be the same, or
related). The model was further analyzed in \cite{Kiers:2002cz}, with emphasis on CP violation properties.

We summarize the left-right model  below without assuming either manifest
 or pseudo-manifest LR symmetry, and with no assumptions about coupling
constants or neutrino masses. We allow a completely arbitrary right-handed quark
mixing matrix $V_{CKM}^R$.

We assume a generic left-right symmetric model based on the gauge group
$SU(3)_C \times SU(2)_L \times SU(2)_R \times U(1)_{B-L}$. The matter
fields of this model consist of three families of quark and lepton fields
 with the following transformations under the gauge group:
\begin{eqnarray}
Q_L^i &=&\left( \begin{array}{c} u_L^i \\ d_L^i \end{array} \right)  \sim \left ( 3, 2, 1, 1/3 \right ),~~
Q_R^i =\left( \begin{array}{c} u_R^i \\ d_R^i \end{array} \right) \sim \left ( 3,1, 2, 1/3 \right ),\nonumber \\
L_L^i& = &\left( \begin{array}{c} \nu_L^i \\ e_L^i \end{array} \right) \sim\left( 1,2, 1, -1 \right),~~
L_R^i= \left( \begin{array}{c} \nu_R^i \\ e_R^i \end{array} \right) \sim \left ( 1,1, 2, -1 \right ),
\end{eqnarray}
where the numbers in the brackets represent the quantum numbers under \\
$(SU(3)_C , SU(2)_L , SU(2)_R, U(1)_{B-L})$. The Higgs
sector consists of one bidoublet :
\begin{eqnarray}
\displaystyle
\Phi&&= \left( \begin{array}{cc} \phi^0_{1} &\phi^+_{2} \\ \phi^-_{1} &
 \phi^0_{2} \end{array} \right) \sim \left (1,2,2,0 \right),
\end{eqnarray}
with vevs
\begin{eqnarray}
\displaystyle
<\Phi>&&= \left( \begin{array}{cc} k &0 \\ 0 & k^\prime\end{array} \right) .
\end{eqnarray}
Additional Higgs multiplets are needed to break the symmetry to the SM
 and to generate a large mass of $W_R$ compared to $W_L$. One can introduce
doublets
\begin{eqnarray}
\delta_L&=&\left( \begin{array}{c} \delta_L^+ \\ \delta_L^0 \end{array} \right)  \sim \left ( 0, 2, 0,1 \right ), ~~~\delta_R=\left( \begin{array}{c} \delta_R^+ \\ \delta_R^0 \end{array} \right)  \sim \left ( 0, 0, 2,1 \right ) 
\end{eqnarray}
with vevs $<\delta_{L,R}^0>=v_{\delta_{L,R}}$.  If  $ v_{\delta_R}\gg (k, k^\prime, v_{\delta_L})$, this choice can generate a large mass for right-handed gauge boson ($M_{W_R}$) and a large right-handed Dirac neutrino mass.
 A popular alternative is to introduce Higgs triplets
\begin{eqnarray}
\Delta_{L} = \left(\begin{array}{cc}
\frac {\Delta_L^+}{\sqrt{2}}&\Delta_L^{++}\\
\Delta_{L}^{0}&-\frac{\Delta_L^+}{\sqrt{2}}
\end{array}\right) \sim (1,3,1,2),~~~
\Delta_{R}  =
\left(\begin{array}{cc}
\frac {\Delta_R^+}{\sqrt{2}}&\Delta_R^{++}\\
\Delta_{R}^{0}&-\frac{\Delta_R^+}{\sqrt{2}}
\end{array}\right) \sim (1,1,3,2).
\end{eqnarray}
The vev for right-handed triplet Higgs boson
 $<\Delta_{R}>=v_{\Delta_{R}}$ can also produce a large $M_{W_R}$ mass
 and generate a large Majorana neutrino mass.
 
Note that in general neither $\delta_L$ nor $\Delta_L$ Higgs bosons are
required, unless one imposes left-right symmetry on the theory. $v_{\Delta_{L}}$
can generate a Majorana mass for the left handed neutrino, but must be very
small (neutral current constraints). As we want to  keep the model general, we
keep both triplets and doublet representations. 
 
The vector bosons of $SU(2)_R$, $W_R^{\pm}$ and $W^0_R$ mix with the SM
 vectors in the charged and neutral sectors, respectively. Here we are
interested in the charged boson sector, as observation of a $W_R$ boson would be
a clear signal for LR symmetry, so we present the masses and mixing for the
charged states only.
In general, $W_L$ and $W_R$ will mix to form mass eigenstates $W_1$ and $W_2$
\begin{eqnarray}
W_L&=&W_1\cos \xi -W_2 \sin \xi 
\nonumber \\
W_R&=&e^{i\omega}(W_1\sin \xi +W_2\cos \xi)
\end{eqnarray}
with $\xi$ a mixing angle and $\omega$ a CP violating phase
 \cite{Langacker:1984dp}. If $\xi$ is small, then $W_L$ and $W_R$ approximately
coincide with $W_1$ and $W_2$. The mass matrix for the charged bosons will be
\begin{eqnarray}
M^2_{W}  &=&
\left(\begin{array}{cc}
\frac12g_L^2(|k|^2 +|k^{\prime}|^2+|v_L|^2) & -g_Lg_R k^\prime k^\star\\
-g_Lg_Rk^{\prime\star} k & \frac12 g_R^2(|k|^2 +|k^{\prime}|^2+|v_R|^2)
\end{array}\right)\nonumber \\
&=&
\left(\begin{array}{cc}M_L^2&M_{LR}^2e^{i\omega^{\prime}}\\M_{LR}^2e^{-i\omega^{\prime}}&M_R^2
\end{array}\right)
\end{eqnarray}
with $|v_{L,R}|^2=|v_{\delta_{L,R}}|^2+4|v_{\Delta_{L,R}}|^2$ and
$\omega^{\prime}=Arg(k^\star k^\prime)$.
 The mass eigenvalues are
\begin{equation}
M_{1,2}^2=\frac12 \{  M_L^2 + M_R^2 \mp [ (M_R^2-M_L^2)^2+4 |M_{LR}^2|^2 ]^{1/2} \}
\end{equation}
and the angles are 
\begin{equation}
\tan 2\xi=\frac{\mp 2M_{LR}^2}{M_R^2-M_L^2}, ~~e^{i\omega}=\pm
e^{i\omega^{\prime}}
\end{equation}
where $M_L$ and $M_R$ are the mass parameters associated with $SU(2)_L$ and
$SU(2)_R$ groups, respectively. For $|v_R| \gg (|k|^2, |k^\prime|^2, |v_L|^2)$
the masses become approximately
\begin{eqnarray}
M_1^2&\simeq&\frac12g_L^2(|k|^2 +|k^{\prime}|^2+|v_L|^2), \nonumber\\
M_2^2&\simeq&\frac12g_R^2|v_R|^2
\end{eqnarray}
and 
\begin{equation}
\xi \simeq \pm\frac{g_L}{g_R}\frac{2|k k^\prime|}{|v_R|^2} \simeq \sin 2 \beta\left ( \frac{M_L}{M_R}\right)
\end{equation}
where $\displaystyle \tan \beta=\frac{k}{k^{\prime}}$.

The charged right-handed bosons contribute to the charged current
 for the quarks, which is
\begin{eqnarray}
{\cal L}&=&\frac{g_L}{\sqrt{2}}{\bar u}{_{iL}\gamma _\mu V^L_{CKM\,ij}d_{jL} W^{\mu +}_L +\frac{g_R}{\sqrt{2}}{\bar u}_{iR}}\gamma _\mu V^R_{CKM\,ij}d_{jR} W^{\mu +}_R
\end{eqnarray}
and similarly for the leptons, which are allowed to mix with different
 CKM-type matrices. We adopt the Wolfenstein parametrization for the CKM matrix
$V^L_{CKM}$ \cite{Amsler:2008zz}
\begin{eqnarray}
V_{CKM}^L  =
\left(\begin{array}{ccc}
1-\frac{\lambda^2}{2}&\lambda&A\lambda^3 (\rho-i\eta)\\
-\lambda&1-\frac{\lambda^2}{2}&a\lambda^2\\
A\lambda^3(1-\rho-i\eta)&-A\lambda^2&1
\end{array}\right)~.
\end{eqnarray}
For the right-handed CKM matrix we allow arbitrary mixing between
 the second and third generations, or between the first and third generations.
To simplify the notation, we drop the CKM subscript and, following 
\cite{Langacker:1989xa}, denote the parametrizations as $(A)$ and $(B)$, where
\begin{eqnarray}
\label{scenAB}
V^R_{(A)} = \left(\begin{array}{ccc}
1&0&0\\
0&\cos \alpha&\pm \sin\alpha\\
0& \sin \alpha& \mp \cos \alpha
\end{array}\right) ,~~~
V_{(B)}^R  = \left(\begin{array}{ccc}
0&1&0\\
\cos\alpha &0&\pm \sin \alpha\\
\sin \alpha&0& \mp \cos \alpha
\end{array}\right),
\end{eqnarray}
with $\alpha$ an arbitrary angle $(-\pi/2 \le \alpha \le \pi/2)$.
 In parametrization $(A)$, depending on the values of $\alpha$, the dominant
coupling could be $V^R_{ts}$ while in $(B)$, the dominant coupling could be
$V^R_{td}$. These parametrizations are by no means the most general. The most general right-handed quark mixing matrix would be a CKM-type matrix, but with arbitrary entries. The $(A)$ and $(B)$ parametrizations are regions of the parameter space which allow relaxing the mass limit on $W_R$, and obeying the restrictions on $\Delta m_K$  without fine-tuning.

The form of the CKM matrix in the right-handed quark sector affects
 low energy phenomenology, in particular processes with flavor violation, and
thus restricts the mass $M_{W_R}$ and mixing angle $\xi$. 
Several studies for $W_R$ production \cite{Ma:1992ve} and mass
 constraints \cite{Barenboim:1996nd} exist in the literature, but most of them
are based on either the manifest, or the pseudo-manifest LRSM.  We calculate the
production cross section and decays of the $W_R$ bosons in the LRSM with large
off-diagonal entries in the $V_{CKM}^R$ matrix, and express the results as
functions of these parameters, and include constraints from both Kaon and B
meson sectors. We confine ourselves to a general version of the model and make
no specific assumptions about the nature and mass of the neutrinos. We summarize
these constraints in the next section.

\section{Constraints on Left-Right Symmetric Models from low energy phenomenology}
\label{constraints}

Before proceeding with the evaluation of the $W_R$ production and
 decays, we summarize briefly the constraints on the parameter space of the
left-right model, mostly from flavor violating processes, which are relevant to
the study of $W_R$ phenomenology.

\subsection{$K^0-{\bar K}^0$ mixing}

Restrictions coming from Kaon physics have been analyzed by several authors.
 For the pseudo-manifest left-right symmetric model, evaluation of $\Delta M_K$
and $\epsilon_K$ restricts the right-handed charged boson mass, $M_{W_R} >1.8$
TeV \cite{Hou:1985ur}, while in more recent analyses of the model, where parity
or charge conjugation is chosen to be broken spontaneously, $M_{W_R} >2.5$ TeV
\cite{Maiezza:2010ic}. 
In the Langacker and Sankar parametrization, which we study here,
  the limit is much lower $\displaystyle M_2  \ge 300 \frac{g_R}{g_L},  ~340
\frac{g_R}{g_L},~670 \frac{g_R}{g_L},~ 350 \frac{g_R}{g_L} $ GeV, respectively
for the four parametrizations ($V_{(A)}^R, V_{(B)}^R$,  with both $\pm$
signs in the right CKM elements) \cite{Langacker:1989xa}.

\subsection{$B_d^0-{\bar B}_d^0$ and $B_s^0-{\bar B}_s^0$ mixing}

 For the pseudo-manifest left-right model the $W_R-W_L$ box diagram
 does not have an important effect. However, for Langacker parametrization
$U^R_{(B)}$ the limit is  $\displaystyle M_2  \ge 1384 \frac{g_R}{g_L}$ GeV
\cite{Gilman:1983ce}. The most comprehensive analysis of the constraints coming
from B meson mixing and decays in the Langacker parametrization is presented in
\cite{Silverman:1997fx}. For more up-to-date constraints, including the recent
data, see \cite{Frank:2010qv}.

\subsection{ $b \to s \gamma$}

Previous studies in left-right models originate from \cite{Rizzo:1994aj},
 while other constraints are presented in \cite{Beall:1981zq}. The constraints 
on $ b \to s \gamma$ were thoroughly analyzed within the asymmetric model
studied here, as well as in the manifest model in \cite{Frank:2010qv}. There are
less restrictive than those coming from $B_d^0-{\bar B}_d^0$ mixing, but
complementary to $B_s^0-{\bar B}_s^0$. These constraints depend on several
parameters and are difficult to summarize; however, they are included in the
cross section plots.

\section{Production and Decays}
\label{prod-dec}

In this section we investigate the single production cross section at LHC of
 a $W_R^{\pm}$ boson, $ p p \to t W_R $, and decay branching ratios of the
right-handed $W$ boson in the scenarios in which the right-handed CKM matrix is
$V_{(A)}^R$ (called $U_A$ for simplicity), $V_{(B)}^R$ (called $U_B$ for
simplicity),  as in  (\ref{scenAB}),  and compare the results  to those obtained
in the manifest left-right symmetric model (MLRSM). In the MLRSM,  the CKM
matrices in the left- and right-handed quark sectors are the same, and so are
the coupling constants for $SU(2)_L$ and $SU(2)_R$. The only unknown parameter
is the $W_R$ mass; while in $U_A$ and $U_B$ the production and decay rates are
also functions of $\sin \alpha$, the right-handed CKM parameter, as well as the
ratio $g_R/g_L$ of $SU(2)_L$ and $SU(2)_R$ coupling constants. 
The dominant partonic level Feynman diagrams are shown in Figure 1. The index
 $i$ indicates that we sum over the three generations.
\begin{figure}[htb]
\begin{center}
\hspace*{-0.7cm}
\includegraphics[scale=1]{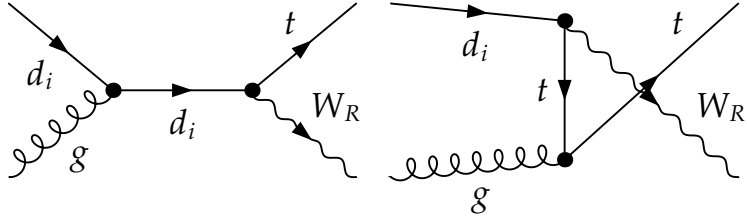}
\end{center}
\vskip -0.3in
      \caption{The $W_R$-top associated production at the LHC}
\label{fig:feyn}
\end{figure}
\begin{figure}[htb]
\begin{center}$
	\begin{array}{ccc}
\hspace*{-1.1cm}
	\includegraphics[width=2.2in,height=2.2in]{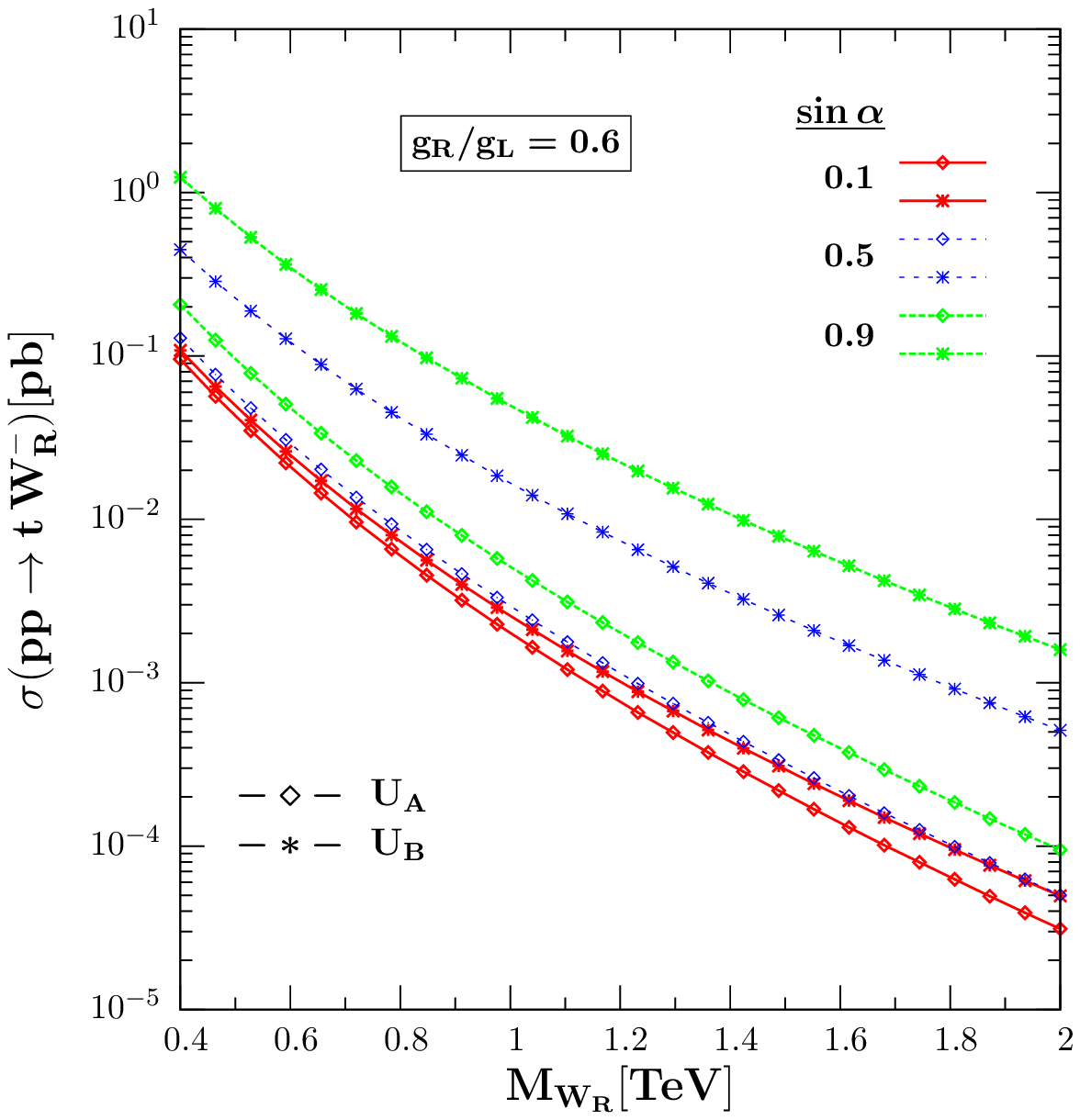}
&\hspace*{-0.2cm}
	\includegraphics[width=2.2in,height=2.2in]{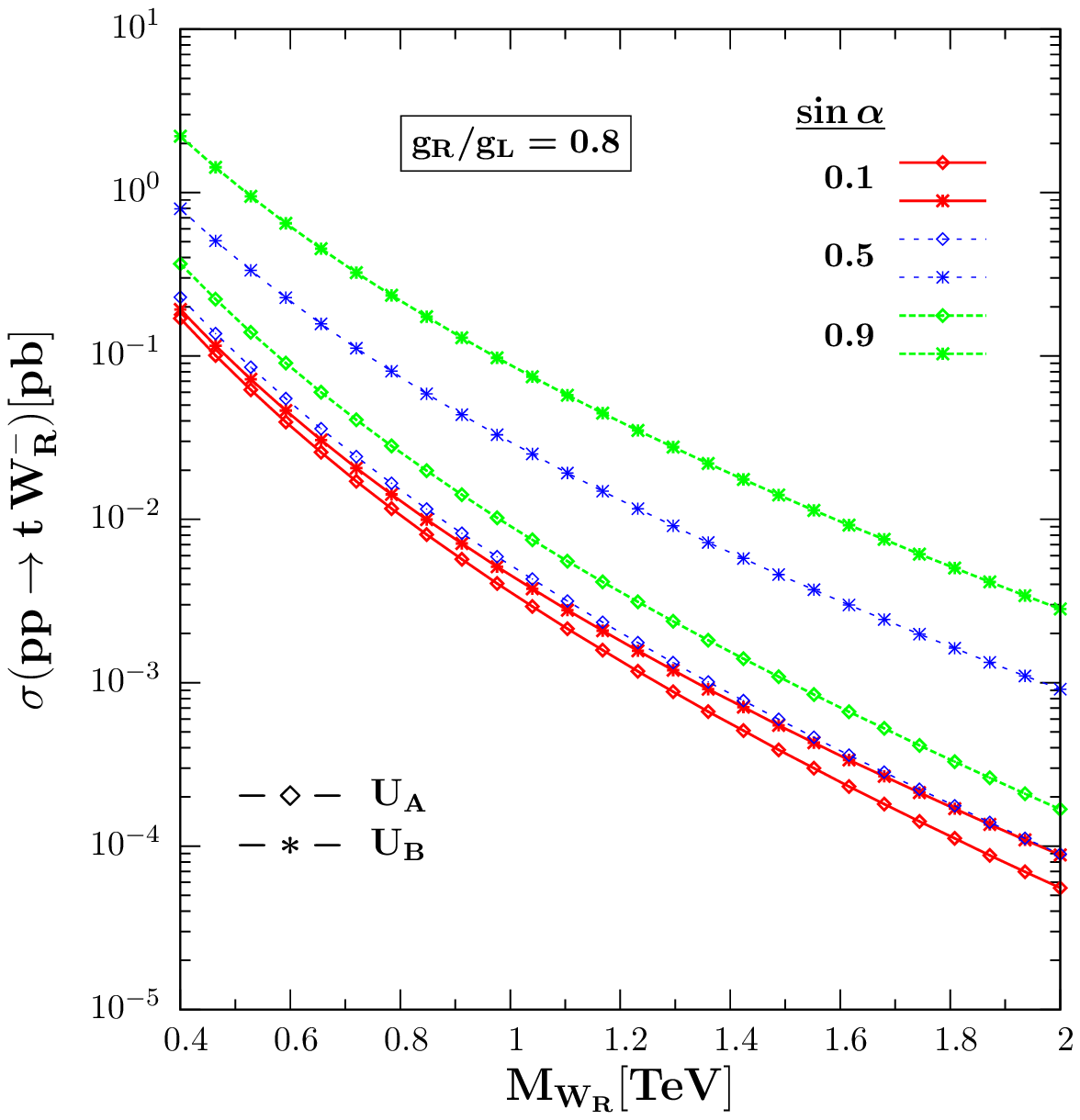}
&\hspace*{-0.2cm}
        \includegraphics[width=2.2in,height=2.2in]{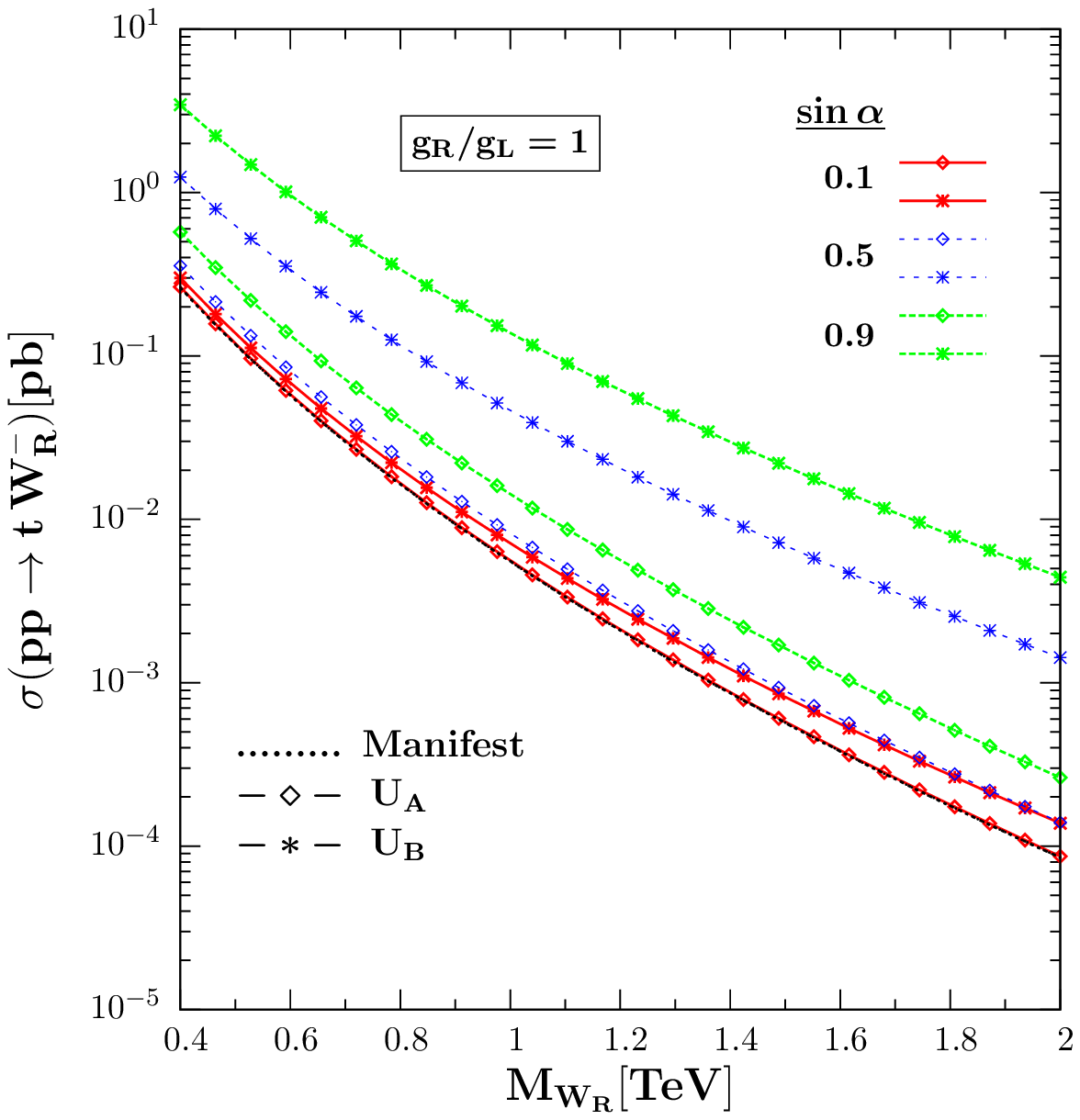} \\
\hspace*{-1.1cm}
	\includegraphics[width=2.2in,height=2.2in]{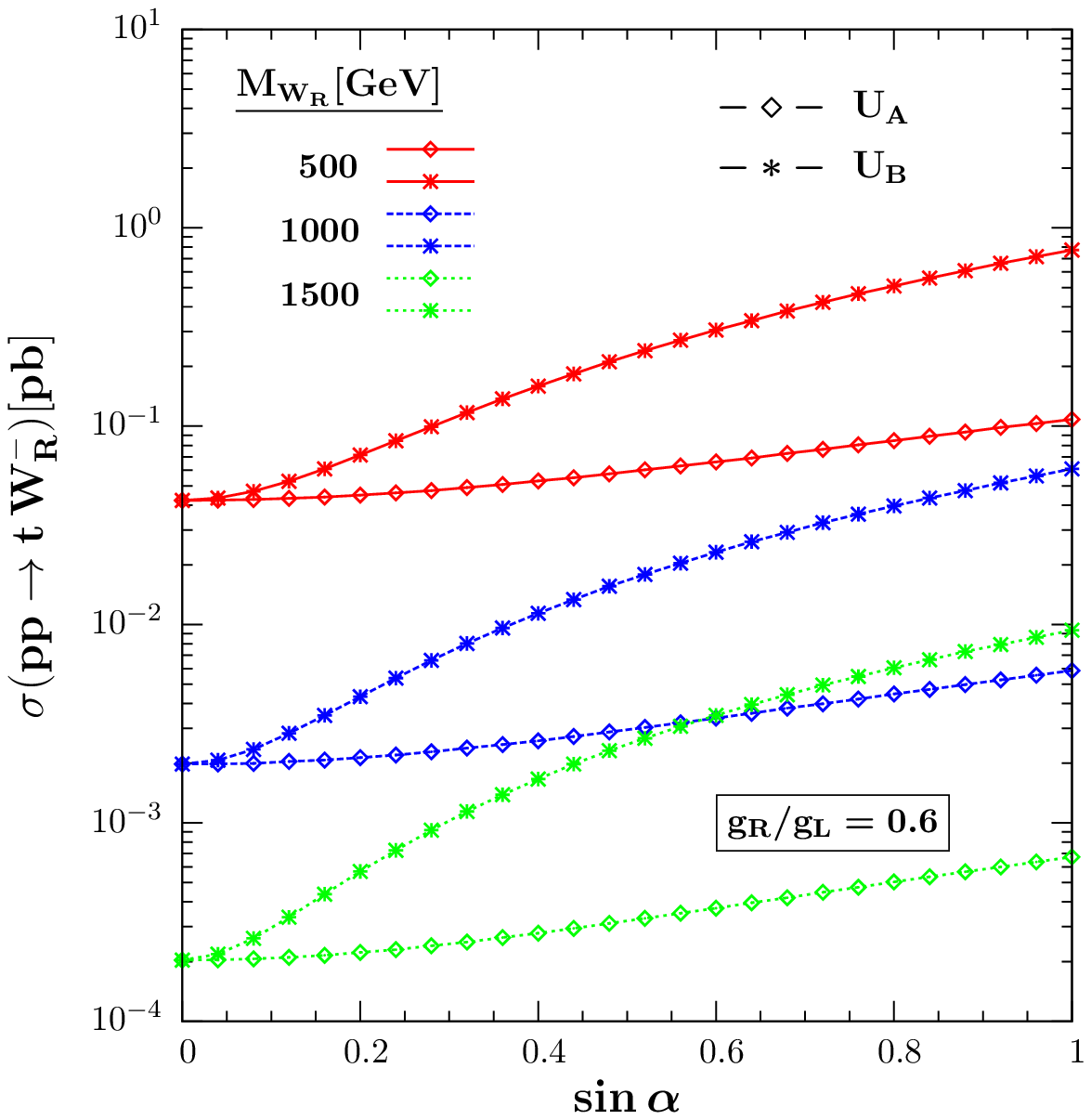}
&\hspace*{-0.2cm}
	\includegraphics[width=2.2in,height=2.2in]{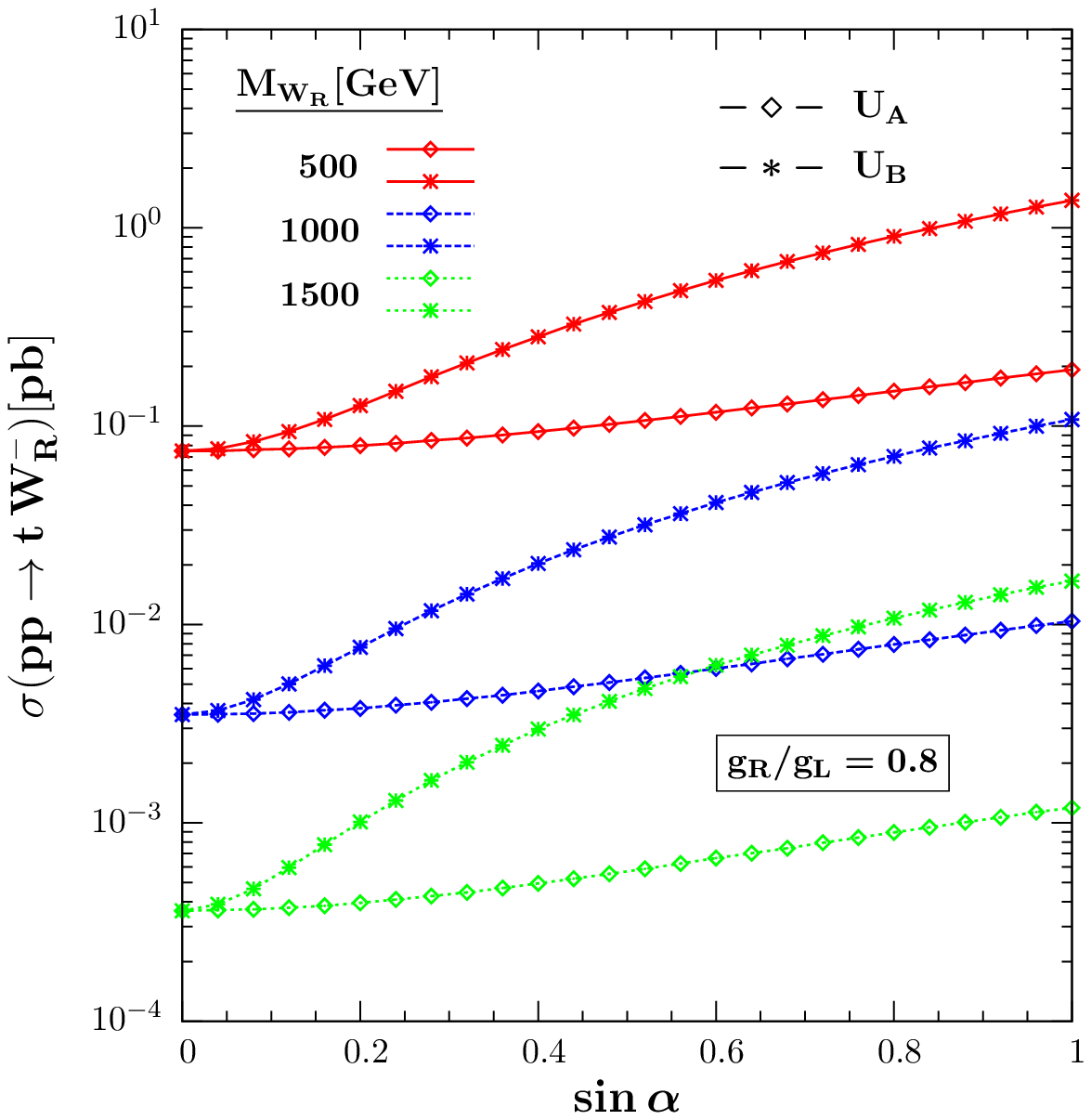}
&\hspace*{-0.2cm}
        \includegraphics[width=2.2in,height=2.2in]{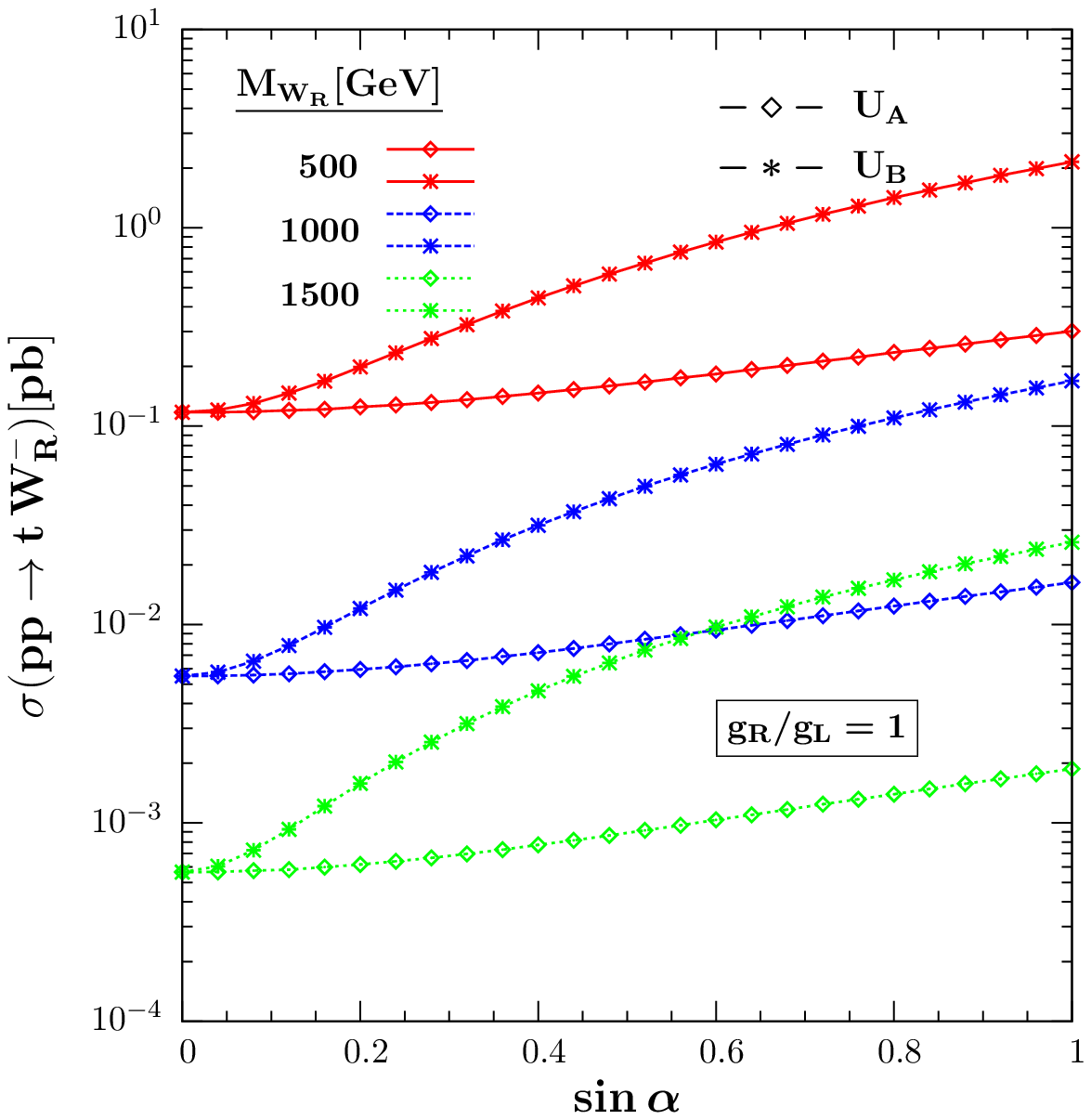} \\
	\end{array}$
\end{center}
\vskip -0.3in
      \caption{$W_R$ production cross-section as a function of $W_R$ mass
 (upper panels) and right-handed CKM matrix parameter $\sin\alpha$ (lower
panels), for the three models described in the text ($U_A, ~U_B$ and manifest
left-right symmetric model).}
\label{fig:sawr}
\end{figure}

In Figure 2, top row we present the single $W_R$ production cross section as
 a function of the $W_R$ mass (in the 400-2000 GeV range) for three values of
$\sin \alpha$.  The three panels correspond to three values allowed for
$g_R/g_L: \,0.6, 0.8 $ and $1$. When $\sin \alpha$ is large, the off diagonal
CKM mixing element $V^R_{td}$ or $V^R_{ts}$ becomes large. As there are more $d$
and $s$ quarks than $b$ in the proton,  this enhances the hadronic contribution
to the cross section for $U_A$ and $U_B$ cases. The production cross section
decreases when $W_R$ mass increases, or $\sin \alpha$ decreases. Similarly, the
production cross section is enhanced by larger $g_R/g_L$. The MLRSM cross
section overlaps with that of model $U_A$ in the case of $\sin\alpha=0.1$
(the right panel in the top row). 

In the bottom row of Figure 2, we explore the dependence of the cross section
 in $U_A$ and $U_B$ on $\sin \alpha$ for three values of $M_{W_R}$. The three
panels again represent cross sections for $g_R/g_L= \,0.6, 0.8 $ and $1$. 
Figure 2 shows that in the region of large $\sin \alpha $ and low $M_{W_R}$ we
can expect large enhancements in the production cross section. For suitable
choices of $\sin \alpha $ and $M_{W_R}$ (light $W_R$ mass and large $\sin\alpha$
region),  the cross section can reach 1 pb or more. The slight difference
between $U_A$ and $U_B$ cross sections is attributed to the relative abundance
of $d$ over $s$ quarks in the proton.
\begin{figure}[htb]
\begin{center}$
	\begin{array}{ccc}
\hspace*{-0.7cm}
	\includegraphics[width=2.2in,height=2.2in]{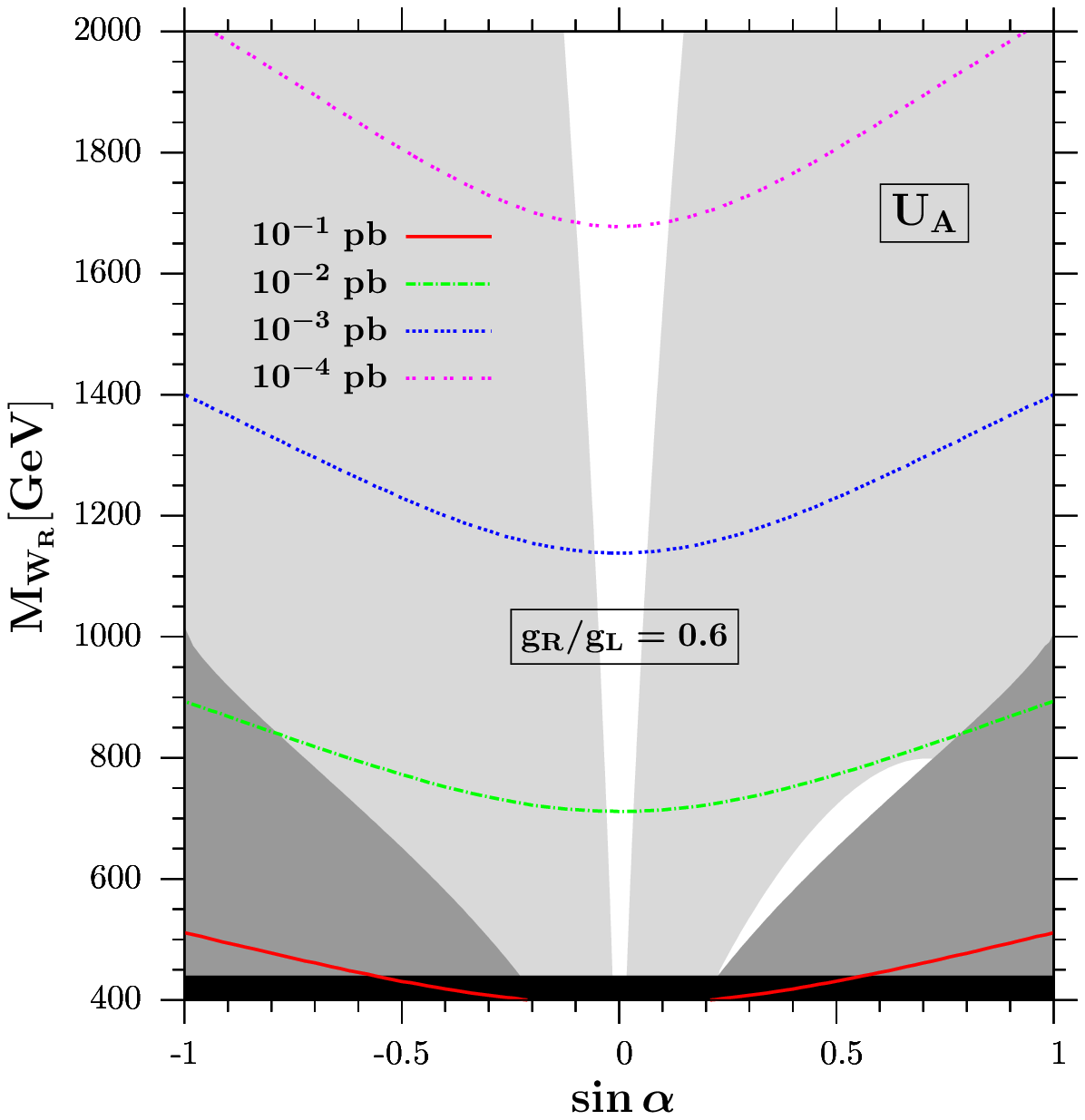}
&\hspace*{-0.2cm}
	\includegraphics[width=2.2in,height=2.2in]{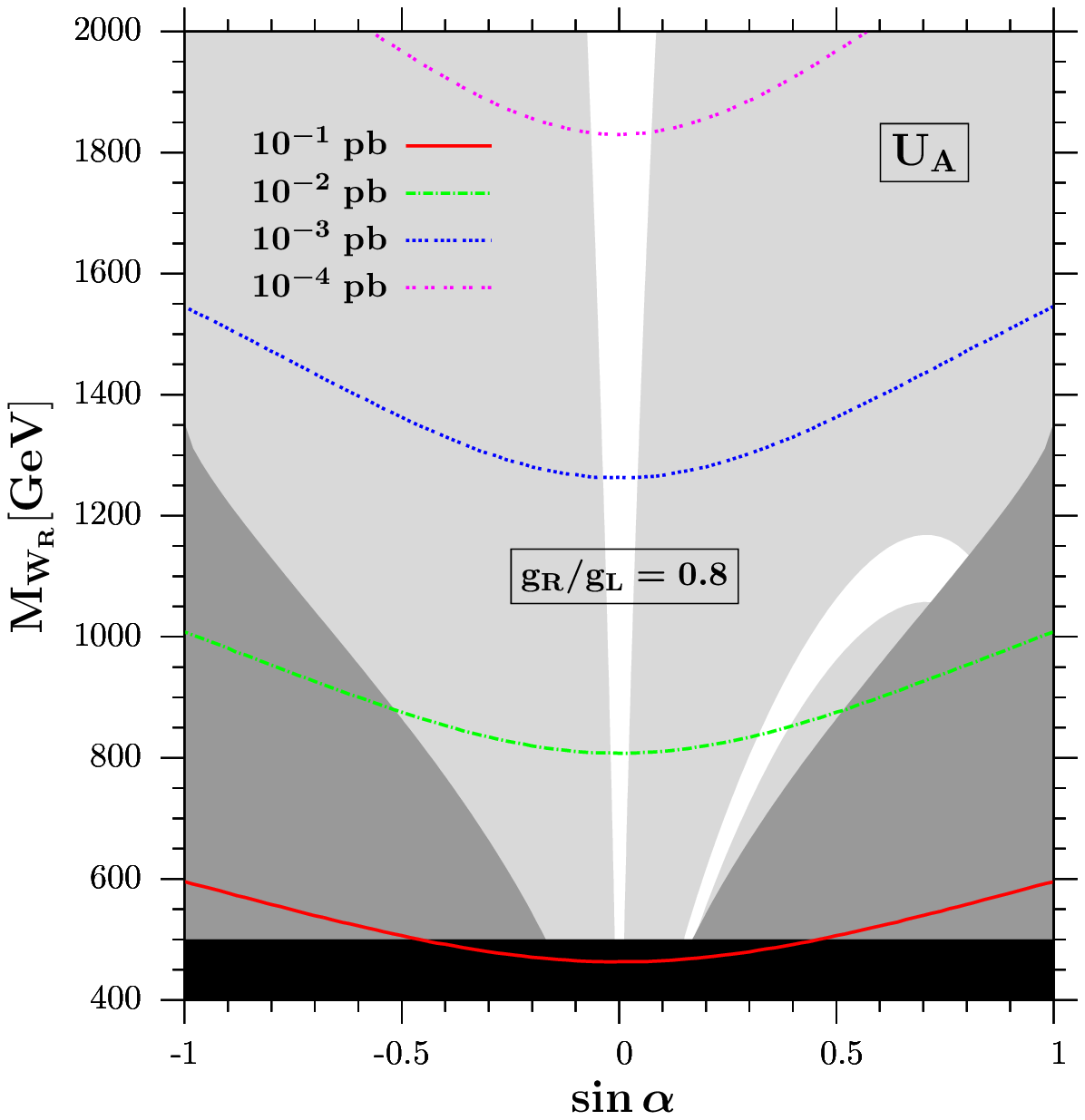}
&\hspace*{-0.2cm}
        \includegraphics[width=2.2in,height=2.2in]{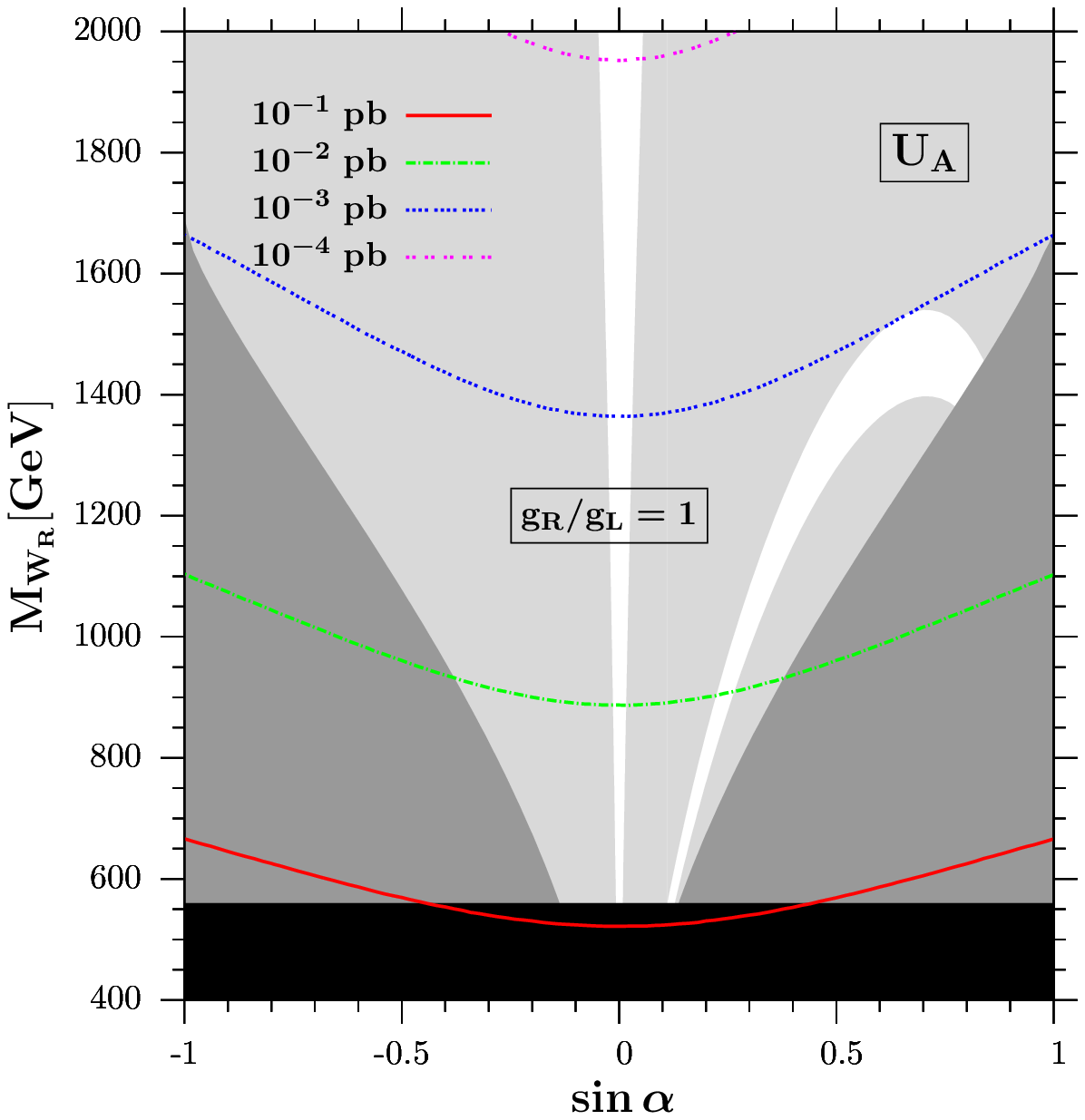} \\
\hspace*{-0.7cm}
	\includegraphics[width=2.2in,height=2.2in]{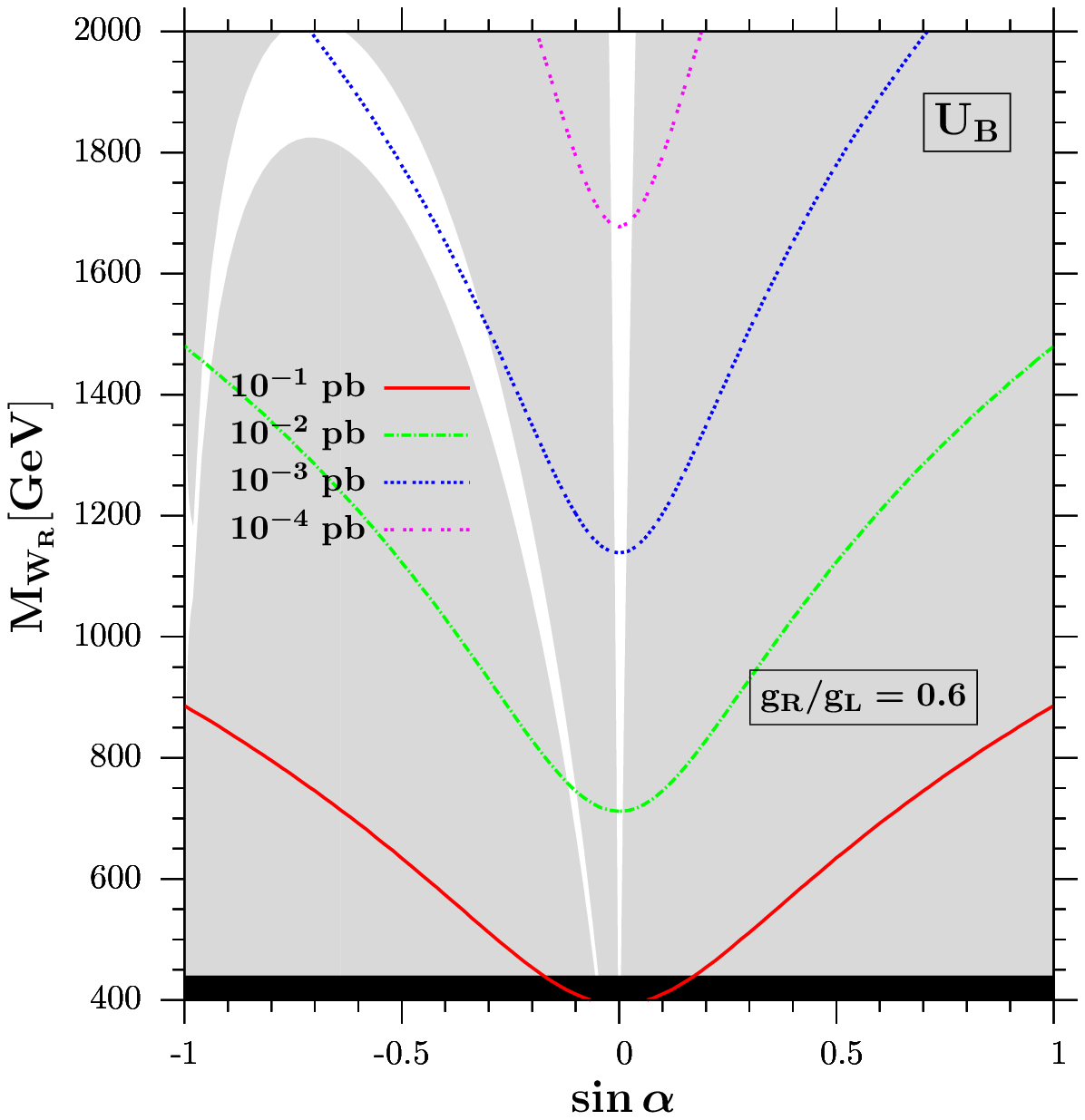}
&\hspace*{-0.2cm}
	\includegraphics[width=2.2in,height=2.2in]{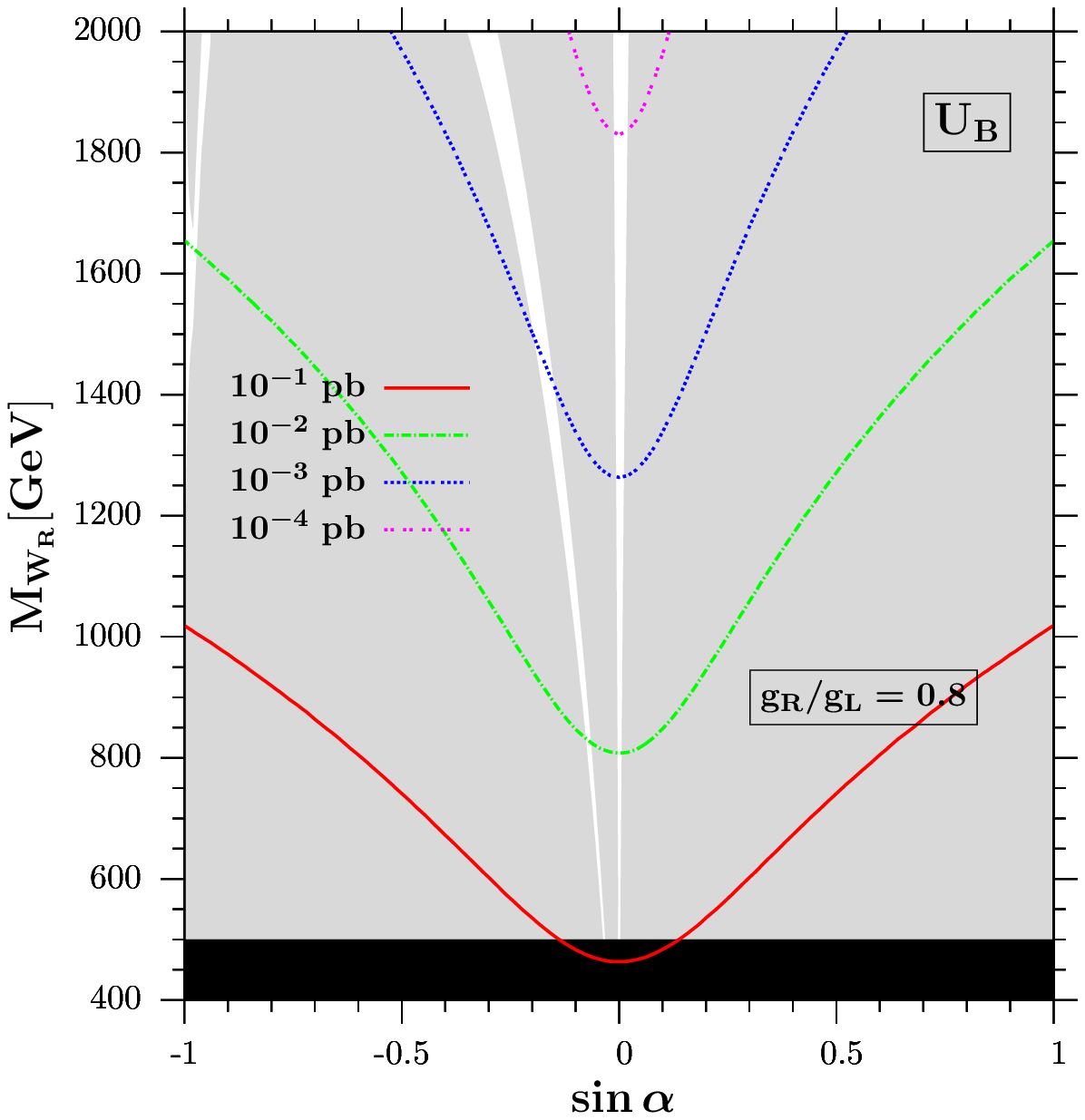}
&\hspace*{-0.2cm}
        \includegraphics[width=2.2in,height=2.2in]{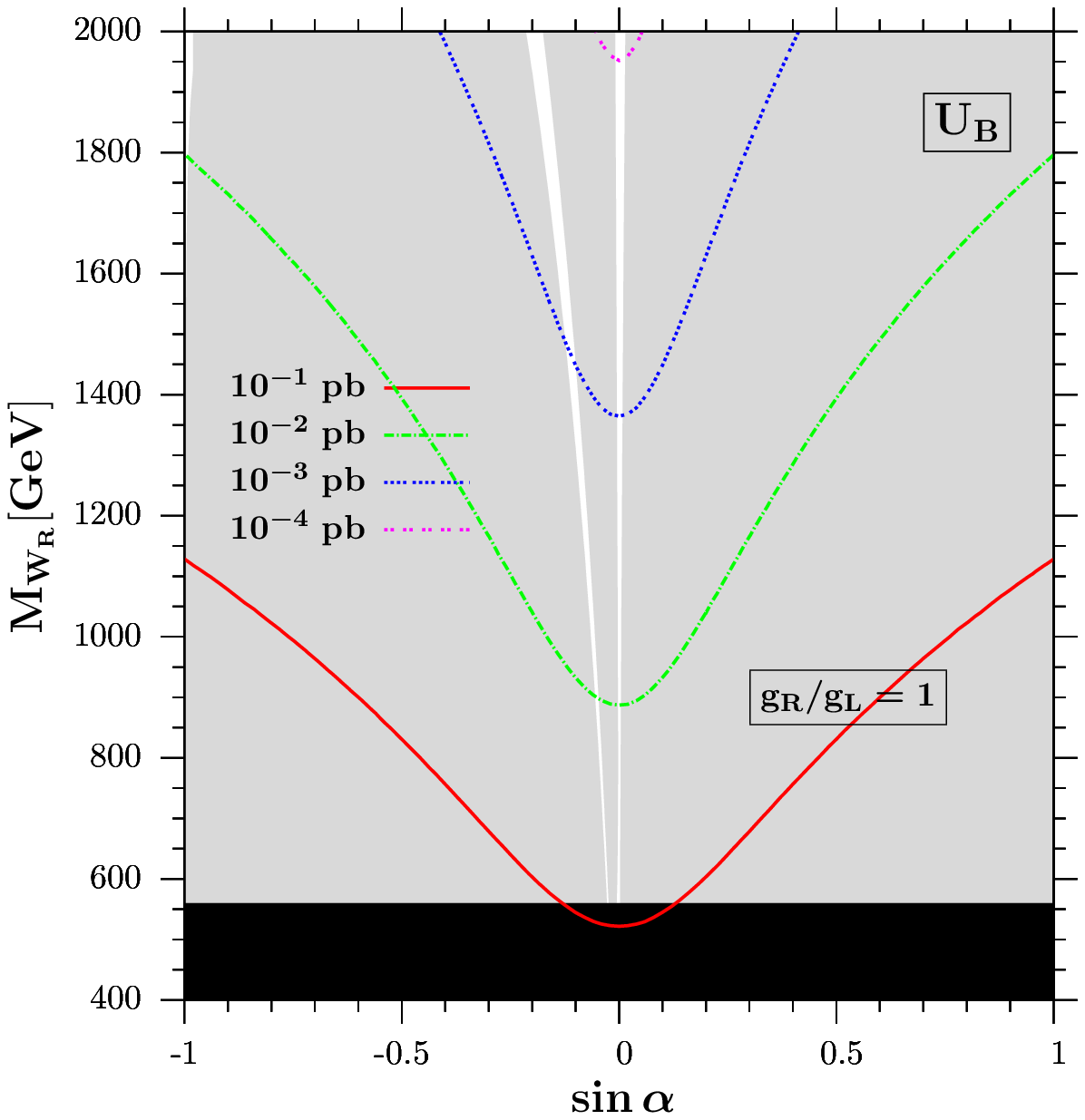} \\
	\end{array}$
\end{center}
\vskip -0.3in
      \caption{ Contour plot of ($M_{W_R}$ vs $\sin\alpha$). Upper row is for
 $U_A$ parametrization in which the $W_R$ production
cross-sections are constrained by both $b\rightarrow s \gamma$ and
$B_s^0-\bar{B}_s^0$ processes. Yellow
shaded regions are excluded from $b\rightarrow s \gamma$ and dashed shaded
regions from $B_s^0-\bar{B}_s^0$.
And dark shaded region indicates the exclusion by L-R mixing angle violation
($\xi < 3 \times 10^{-3}$). Lower 
row is for $U_B$ parametrization where  only $B_d^0-\bar{B}_d^0$ mixing
constrains the production 
cross-section. In both parametrizations we take $M_{H^+}=20$ TeV and
$\tan\beta=30$.}
\label{fig:contour}
\end{figure}

In  Figure 3, we give a contour plot in the $M_{W_R}-\sin \alpha$ parameter
space,
 including constraints from $b\rightarrow s \gamma$,  $B_d^0-\bar{B}_d^0$, and
$B_s^0-\bar{B}_s^0$ processes. This plot correlates restrictions on $\sin
\alpha, M_{W_R}$,   $g_R/g_L$ and production cross sections.  In the top row, we
show the plot for the $U_A$ parametrization. This parametrization is constrained
by  $b\rightarrow s \gamma$ branching ratio (in yellow) and and
$B_s^0-\bar{B}_s^0$ mixing (dashed). The three panels represent increasing
values of coupling constants ratio $g_R/g_L= 0.6, 0.8$ and $1$. The dark shaded
parameter region at the bottom (increasing with larger $g_R/g_L$) represents
restrictions due to the $W_L-W_R$ mixing angle $\xi < 3 \times 10^{-3}$. The most stringent phenomenological inputs which restrict the $W_L-W_R$ mixing angle $\xi$ are: weak universality for light neutrinos, partial conservation of axial-vector-current in $K \to 2\pi$ and $K \to 3 \pi$  and constraints on $W_L$ mass, which is reduced by increasing $\xi$ \cite{Langacker:1989xa}. The
parameter space is overall very restricted. For smaller $g_R/g_L$ there is a
stable allowed region around $\sin \alpha=0$, which is decreasing with
increasing $g_R/g_L$. However, for all coupling ratios, there is a parameter
space  allowed, where $\sin \alpha $ is large and positive, and the $W_R$ mass
can relatively light ($M_{W_R}=600-700$ GeV for $g_R/g_L=0.6$) or intermediate
($M_{W_R}=1400-1500$ GeV for $g_R/g_L=1$). For these cases the cross section can
be of order $10^{-2}$ pb.

The bottom row of Figure 3 presents the same restrictions on the $M_{W_R}
-\sin \alpha$ parameter space in the $U_B$ parametrization. The three panels
again represent restrictions for $g_R/g_L= \,0.6, 0.8 $ and $1$. The
restrictions come from $B_d^0-\bar{B}_d^0$ (shaded) and the $W_L-W_R$ mixing
angle $\xi < 3 \times 10^{-3}$ (dark shaded--this constraint is the same as in
the upper row). The $U_B$ parametrization is much more restricted, reflecting
the stringent restrictions from $B_d^0-\bar{B}_d^0$ mixing. While the same
region around $\sin \alpha=0$ exists in all graphs, it is shrunken very close to
zero, especially for $g_R/g_L=1$. The region for $\sin \alpha$ away from zero
(in this case negative) is significant only for $g_R/g_L=0.6$ and larger values
of the $W_R$ mass. Still, there is a small parameter space available for
$M_{W_R}=1.8-2$ TeV. But the cross section expected in this region is of order
of $10^{-3}$ pb, smaller by a factor of 10 than that for the $U_A$
parametrization.

\begin{figure}[htb]
\hspace{-.7in}
$\begin{array}{cc}
 \includegraphics[width=3.3in]{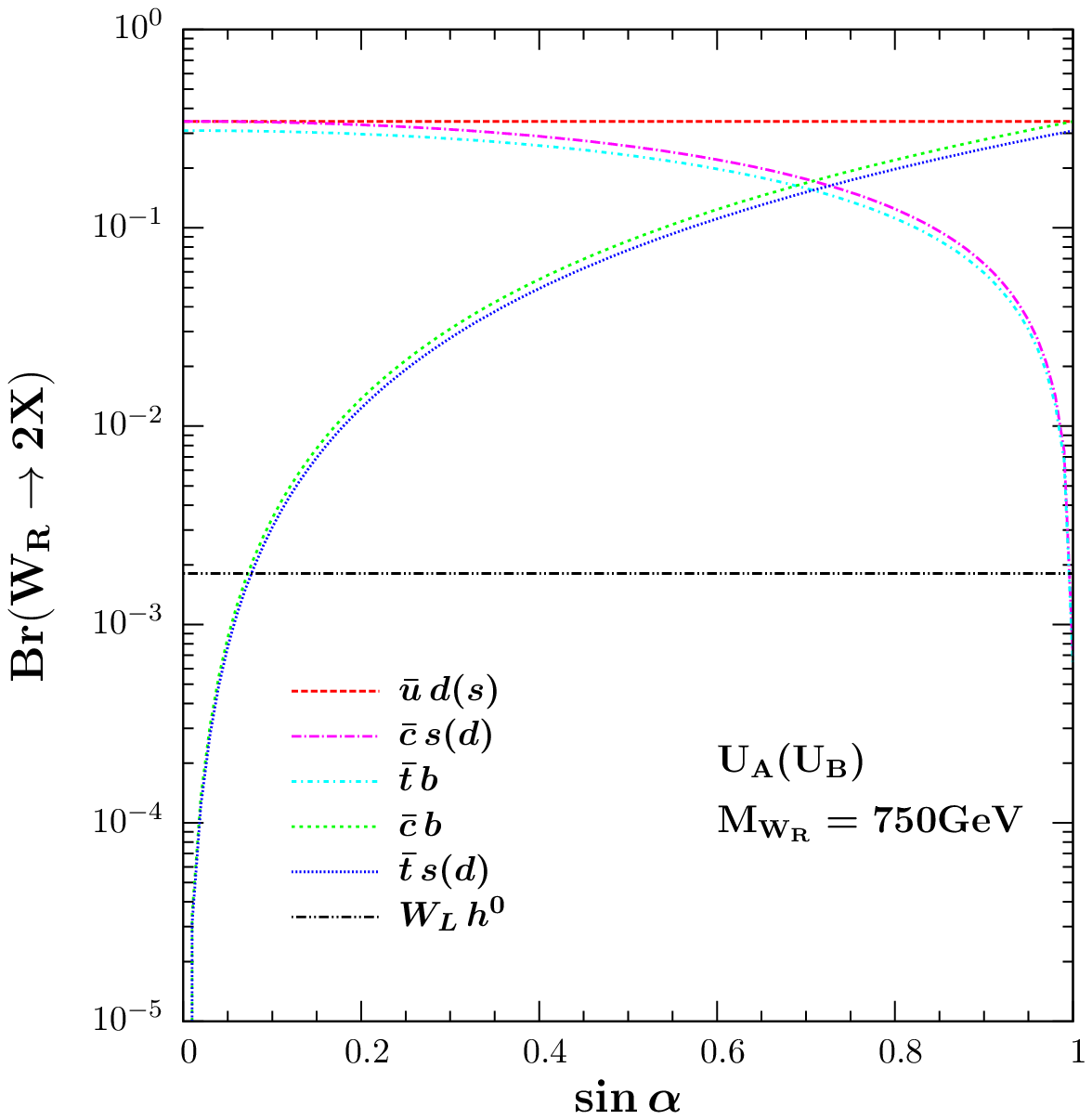} &
 \includegraphics[width=3.3in]{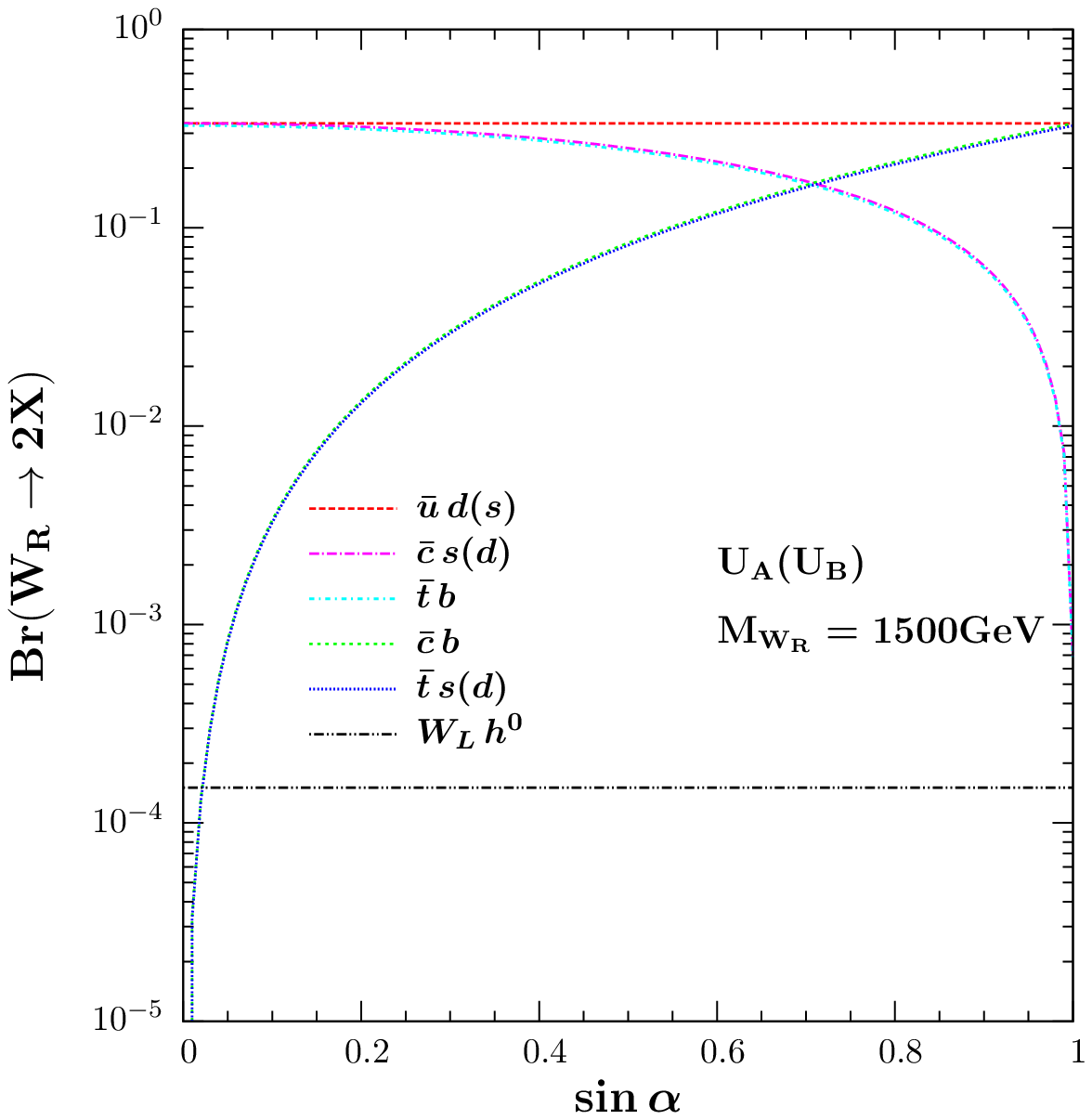} \\
 \includegraphics[width=3.4in]{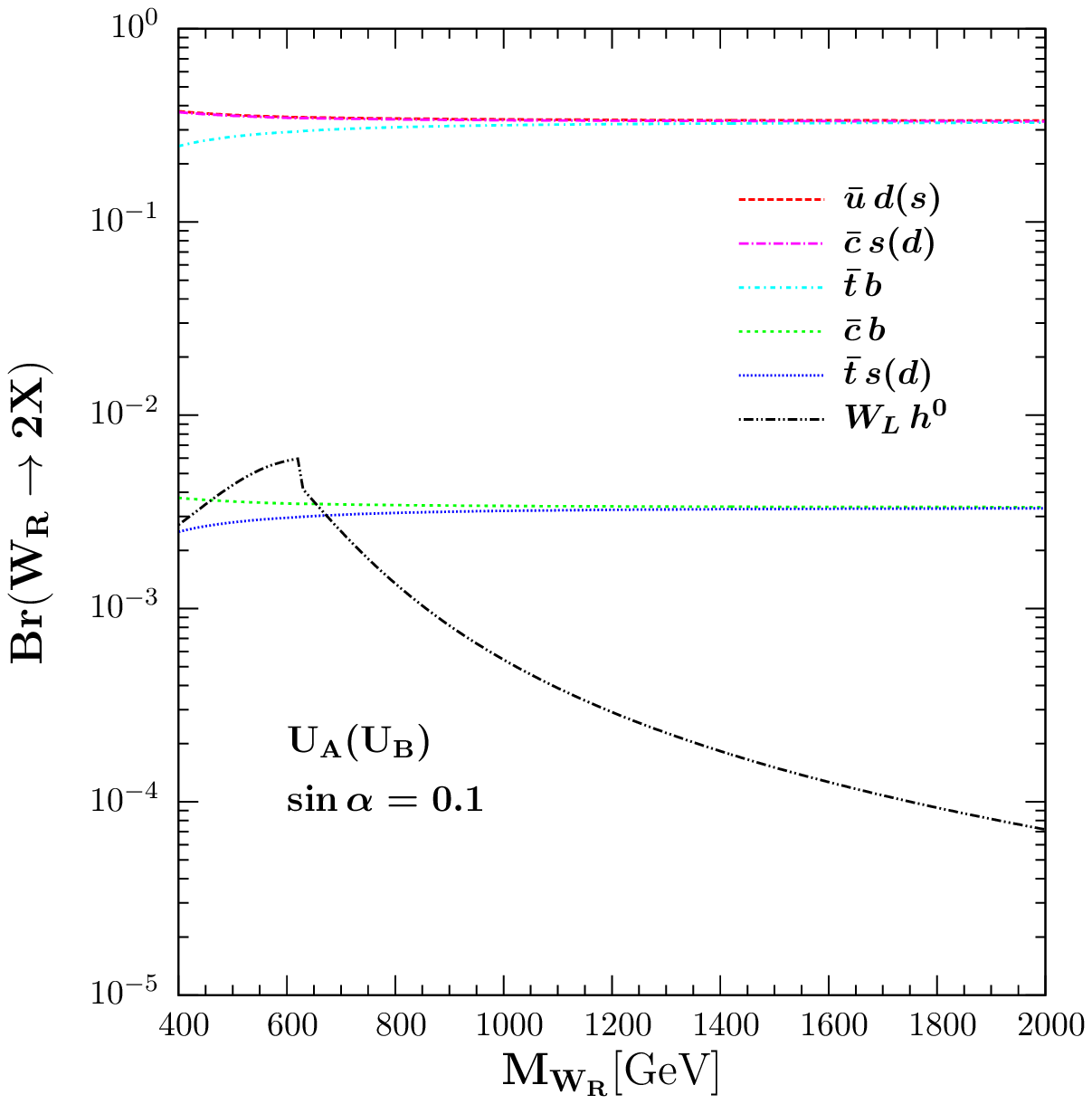} &
 \includegraphics[width=3.4in]{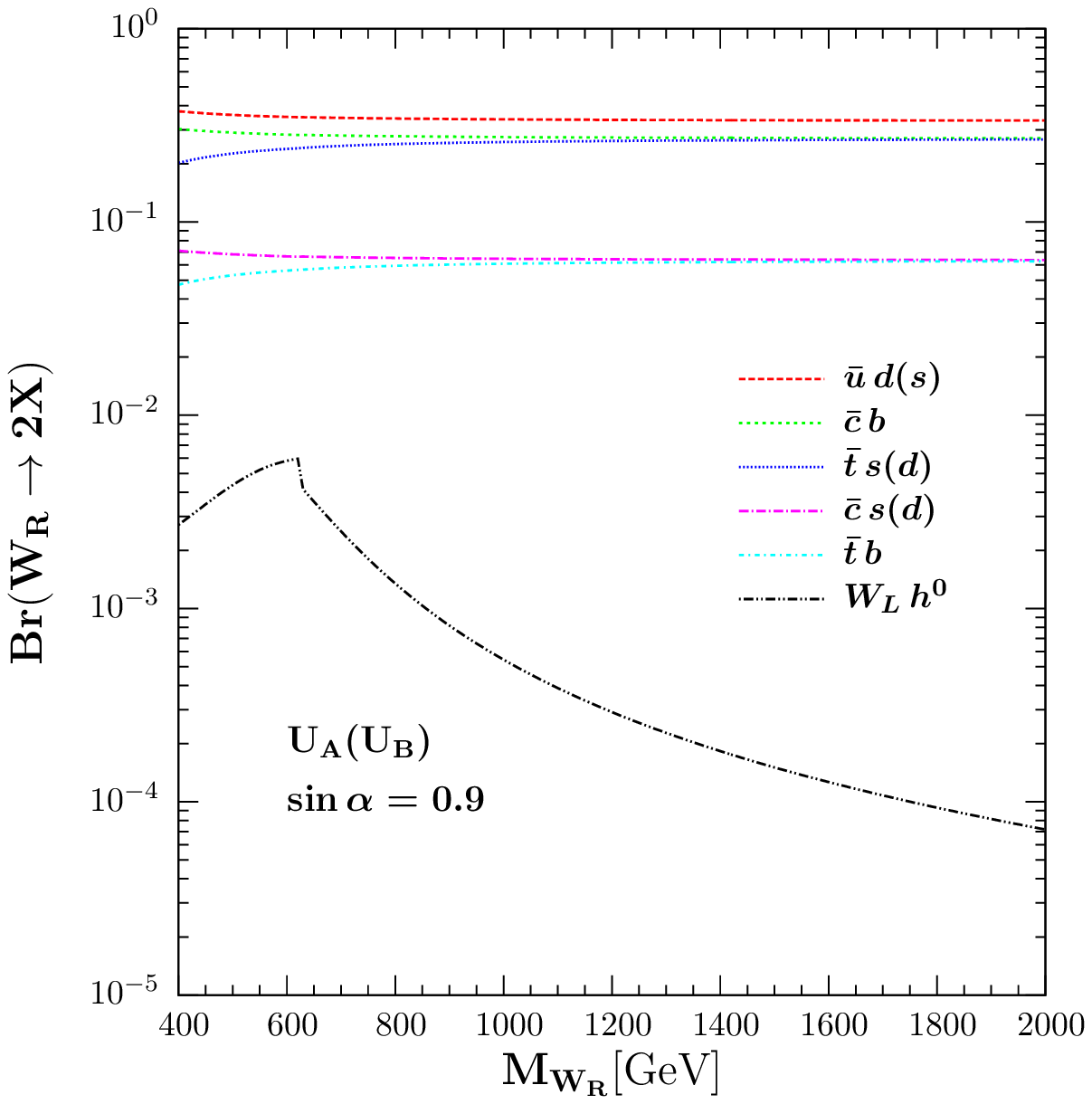}
\end{array}$
\caption{Branching ratios of $W_R$ decays  as functions of $\sin\alpha$
 (above panel) and $W_R$ 
mass (below panels). The $W_R$ mass is fixed at 750 GeV in left top and 1500
 GeV in right top panel, while  $\sin\alpha$ is fixed to 0.1 in left bottom
panel and 0.9 in right one.}
\end{figure}

In Figure 4 we present the branching ratios of $W_R$ decays into quarks,
  and a representative one into $W_L h^0$ (assuming this decay has the phase
space required to proceed) in the asymmetric left-right model. In the top
panels, we analyze the decay width into quarks, as  a function of $\sin \alpha$,
for both $U_A$ and $U_B$ scenarios. The left panel corresponds to $M_{W_R}= 750$
GeV, the right one to $M_{W_R}=1.5$ TeV. It is possible to include both
parametrizations in  one plot because, between these two scenarios, the CKM
matrix elements involving $s$ and $d$ quarks mixing with $t$ quarks are
switched, and although the masses of these quarks are not identical, it does not
significantly impact on the branching ratios. While $W_R^- \to d(s) {\bar u}$ is
the dominant decay for both cases, for large $\sin \alpha$ the branching ratios
to $b {\bar c}$ and $d(s) {\bar t}$ become comparable; while for low $\sin
\alpha$ the branching ratios to the same-generation pairs, $b {\bar t}$ and $s
{\bar c}$, are large. The leptonic decays $W_R^- \to l^- {\bar \nu}_R, \,(l=e,
\mu)$ are not presented here, as we wanted to avoid extra assumptions on the
nature of the neutrinos and their masses. Many other decay channels are
possible, but we have chosen to only  illustrate $W_Lh^0$. It is possible that,
for a range of the parameters, there is sufficient phase space for other decays
(to leptons, $h^0 H^\pm$, $Z_L H^\pm$,...) to proceed, but all require further
assumptions. In our analysis, charged Higgs and all other neutral Higgs bosons
except for $h^0$ are heavy, so these channels are not open. The branching ratio
to $W_Lh^0$  is independent of $\sin \alpha$ and always dominated by branching
ratios to quarks. 

The panels in the bottom row show the dependence on the same branching
 ratios as a function of $M_{W_R}$, for $\sin \alpha=0.1$ (left panel) and $\sin
\alpha=0.9$ (right panel). The dominance of the $d(s) {\bar u}$ decay mode
persists, and is independent of $\sin \alpha$, a consequence of the form chosen
for $V^R_{CKM}$ to agree with Kaon phenomenology. The branching ratios are
independent of the mass of the $W_R$, with the exception of $W_Lh^0$. Note that
the branching ratios also do not depend on the coupling constant for $SU(2)_R$
(or $g_R/g_L$), as it appears as an overall factor in both the partial decay
width and total width formulae.
\begin{figure}[htb]
\hspace{-0.8in}
 \includegraphics[width=3.3in]{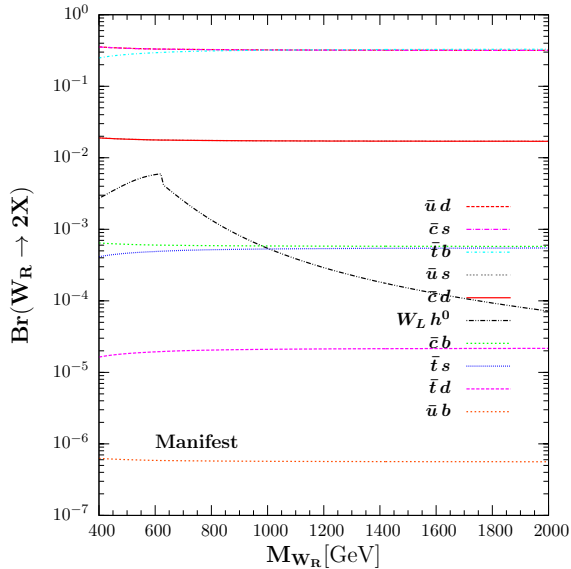}
\caption{Branching ratios of $W_R$ decays as functions of $W_R$ mass
 in Manifest Left-Right Model.}
\end{figure}

Figure 5 illustrates the branching ratios of all decay modes of the $W_R$ boson
  in the MLRSM. The main difference from the point of view of observability is
that in scenarios $U_A, \, U_B$ there are 5 $q {\bar q}^\prime$ decay modes with
branching ratios between $5 \times 10^{-2} - 2 \times 10^{-1}$, while in MLRSM
there are only 3 (for $M_{W_R} >500$ GeV). In both cases, all the other
branching ratios are much smaller and very similar in all three scenarios. For
the purpose of explicit branching ratio calculations, we considered the case in
which the bi-doublet Higgs boson is supplemented by triplet Higgs bosons. Under
this assumption we diagonalized the Higgs mass matrix and calculated the Feynman
rules. We expect the case with doublet Higgs bosons to yield very similar
results when we impose experimental constraints.

\section{Signal and background for $W_R$ production at the LHC}

\begin{figure}[htb]
\hspace{-0.8in}
 \includegraphics[width=3.3in]{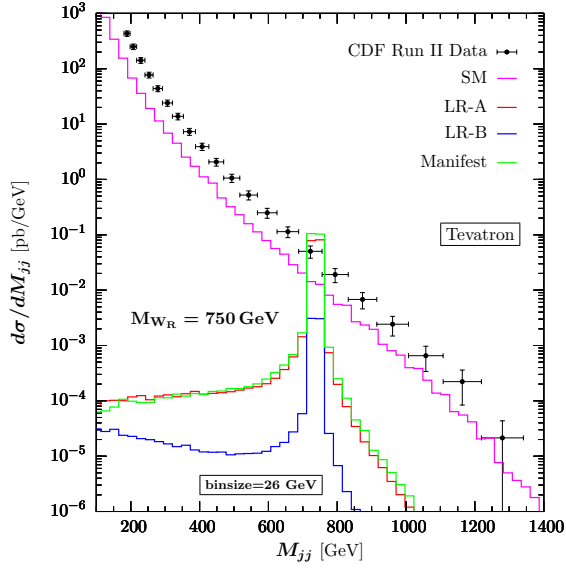}
\caption{Differential cross section for the $dijet$ mass spectrum for  $W_R$ decays in the $U_A$, $U_B$ parametrizations and in the Manifest Left-Right Model, compared to the SM background and the CDF data. 
It is possible to show that the SM curve fits very well with the CDF Run II data after
including NLO perturbative QCD corrections. Our SM curve should be taken as a rough estimation.}
 \label{tevatron}
\end{figure}

Before proceeding with the analysis of the $W_R$ production signal at the LHC, we consider the signal at the Tevatron, from $p \bar{p} \to W_R \to dijet$.  The $dijet$ data is already available  from CDF Run II \cite{CDFdijet}, and the analysis  shows no significant evidence for a narrow resonance. This is used to put mass constraints on several beyond the SM particles, including the $W^\prime$. To compare the data with our model, we used {\tt CALCHEP} software
\cite{Pukhov:2004ca} and implemented the model into the software. To obtain the $dijet$ spectrum we used the following detector cuts at $\sqrt{s}=1.96$ TeV: $p_T>40$ GeV, $|y|<1,~|\eta|<3.6$ and $R_{\rm cone}=0.7$ (jet cone angle). The parameters used to generate Figure \ref{tevatron} are $M_{W_R}=750$ GeV, $g_R/g_L=1$, $\sin \alpha=0.2(-0.05)$ for $U_A(U_B)$. The $dijet$ process is dominated by $s-$channel contributions.  From the figure we see that under these conditions, the $W_R$ signal falls below the CDF data and  would not be observable at the Tevatron. Thus we cannot expect to extract meaningful mass bounds for $W_R$, even for a relatively light gauge boson.

\label{signal-back}
\begin{figure}[b]
 \includegraphics[scale=0.9]{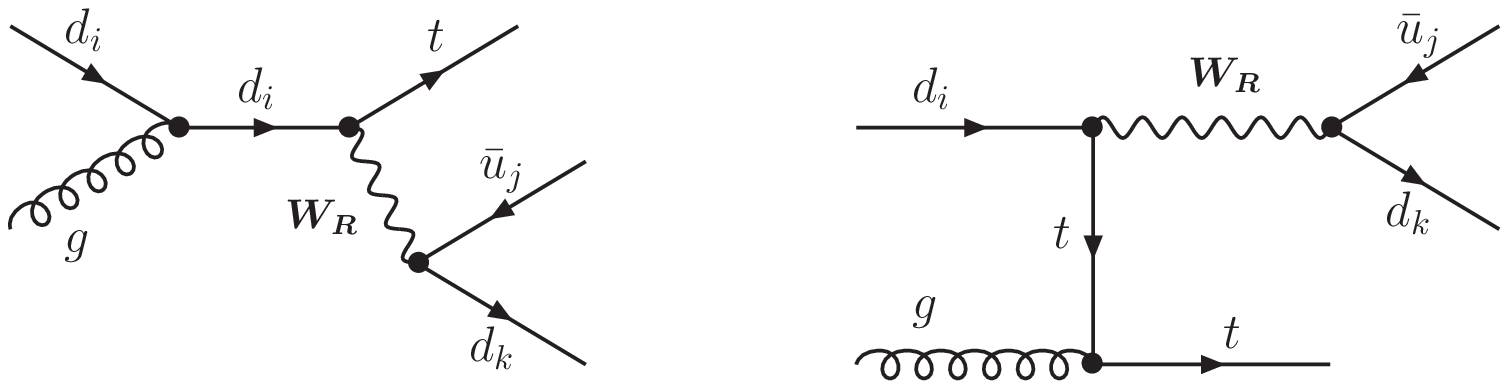}
\caption{The signal $pp \to t \,W_R \to t (jet\, jet)$ }
\end{figure}

We then proceed with the  investigation of the $W_R$ production signal at the LHC.  We simply
 considered a single $top$ production associated with a $dijet$ through a
$W_R$ exchange in both $s$- and $t$-channel processes as in Figure 6. Assuming
$b$-jets are tagged and further $top$ decays are reconstructed, we selected only
light quarks ($u,c,d$ and $s$) in jets. In order to compare our signal with the
background we accounted all the possible $top+dijet$ processes in the SM final
state. For the signal analysis we used again the implementaion of our model into the {\tt CALCHEP} software
\cite{Pukhov:2004ca}. We
also introduced some basic detector cuts on the pseudo-rapidity ($|\eta|<2$) and
on the transverse energy ($p_T > 30$GeV). We assume that in both our model and
in the SM, the top quark will decay as predicted, and it can be reconstituted.
We have chosen $W_R$ decays to quarks, rather than leptons, because we wanted to
avoid assumptions on the nature and masses of neutrinos. Also, jets can be
easily identified and this decay mode does not involve any missing energy,
making it easier to detect a $W_R$ resonance. We also restricted the decay
products to jets (light quarks only) to avoid $t{\bar t}$ production. In the
case of considering $W_R\to \bar{t} d_i$, the SM background would be
$t\bar{t}j$ and could be significant.
\begin{figure}[t]
\vspace{-0.4in}
\hspace{-.5in}
$\begin{array}{cc}
 \includegraphics[scale=0.57]{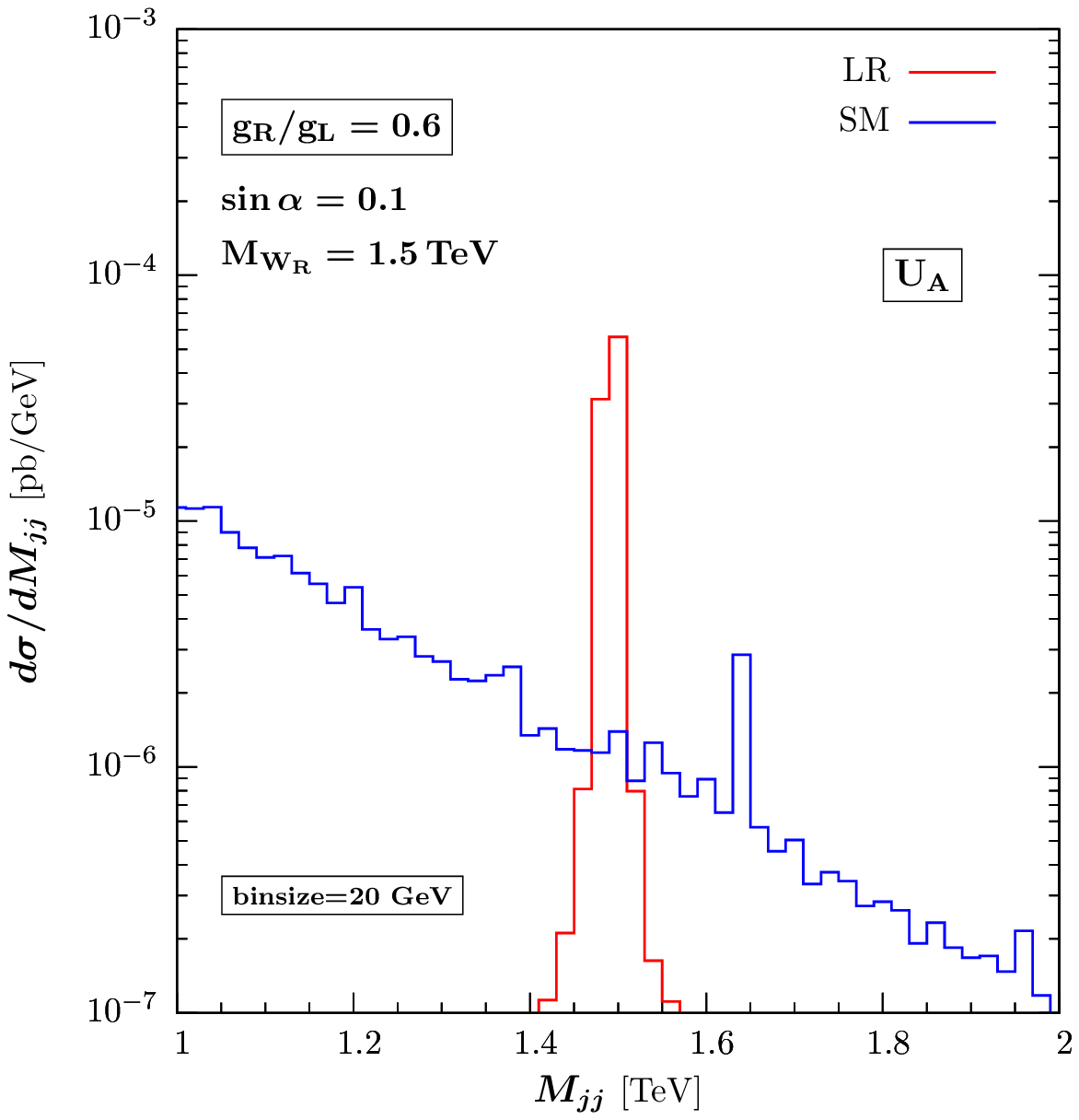} &
 \includegraphics[scale=0.57]{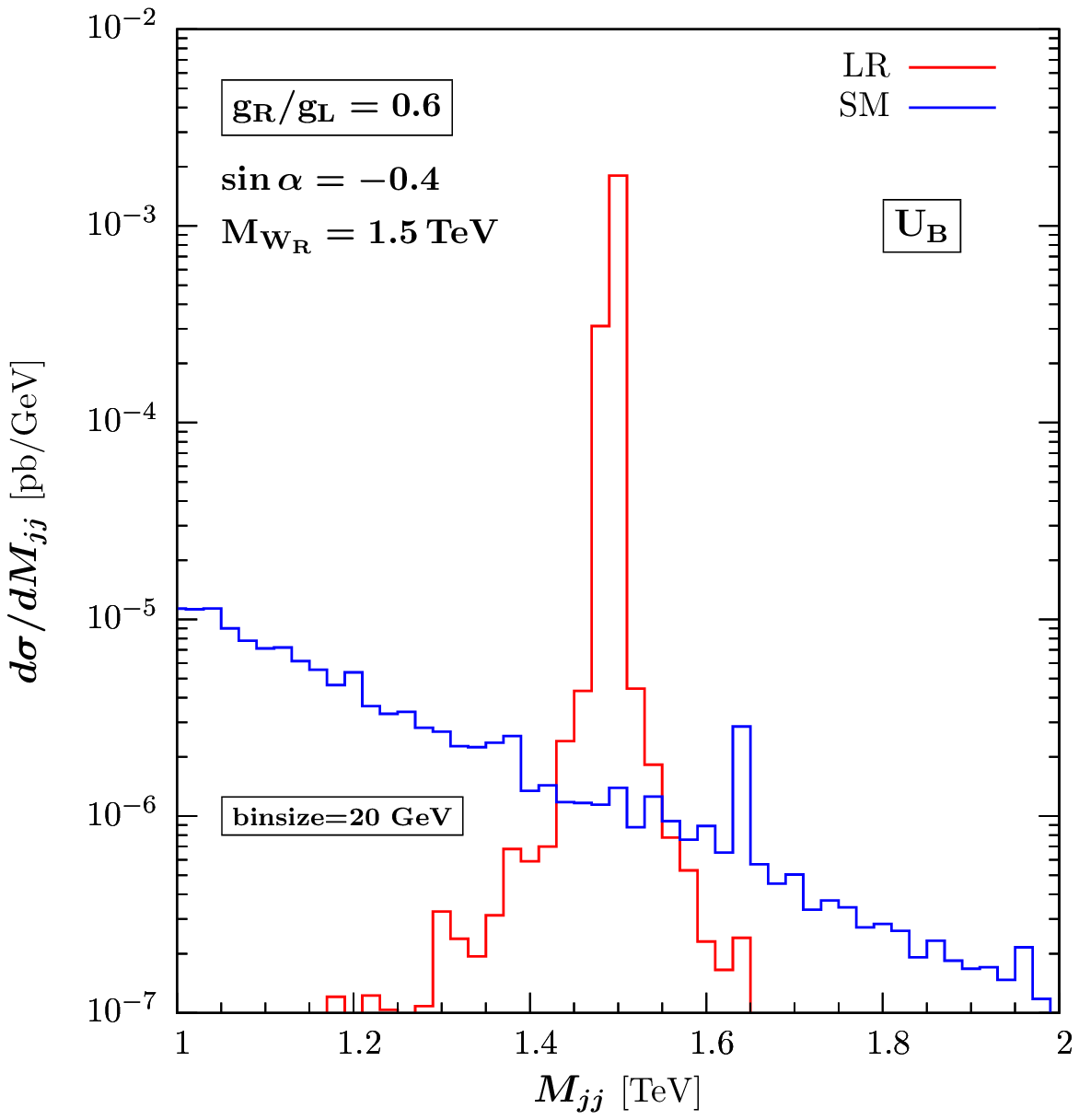} \\
 \includegraphics[scale=0.57]{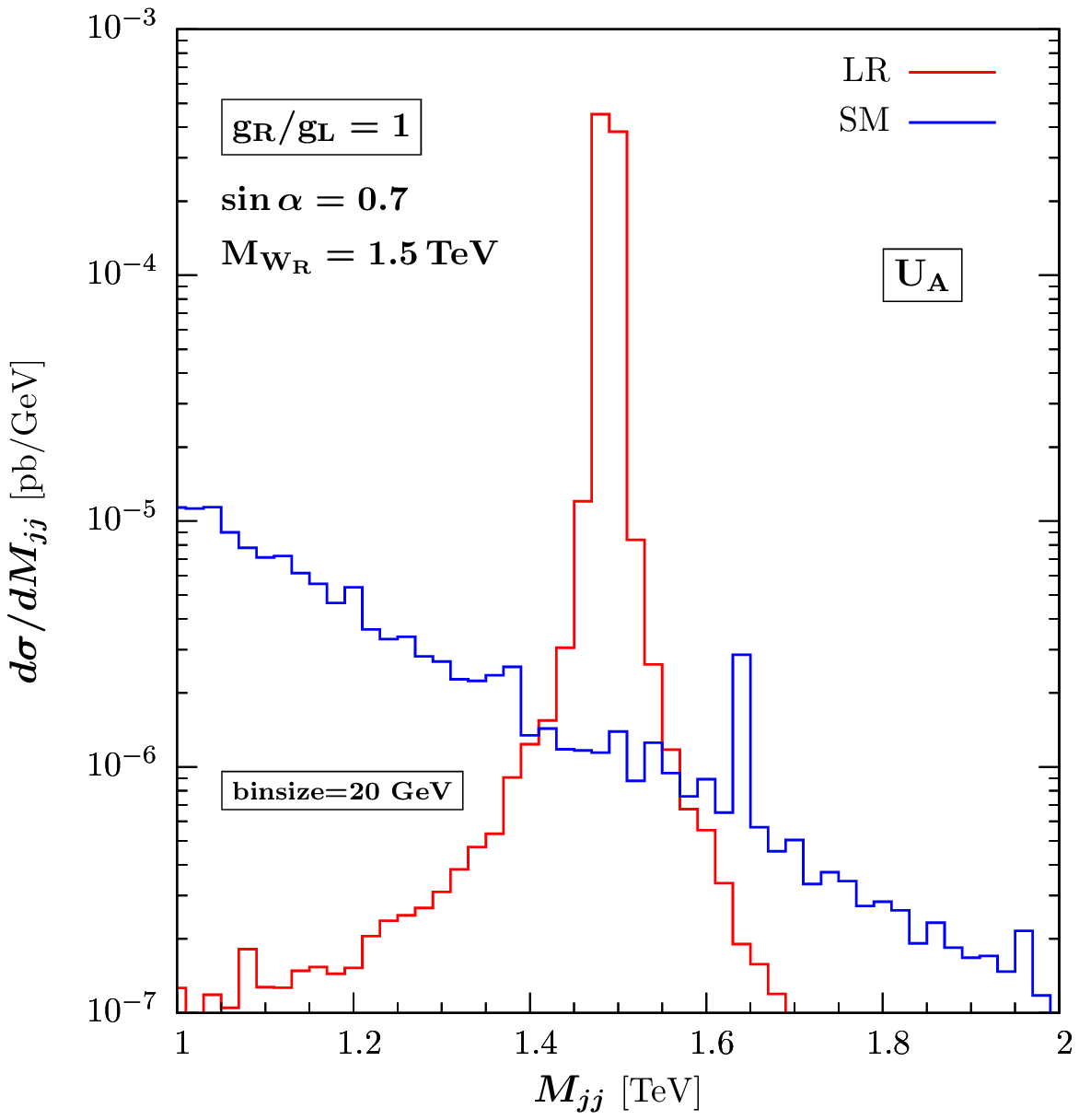} &
 \includegraphics[scale=0.57]{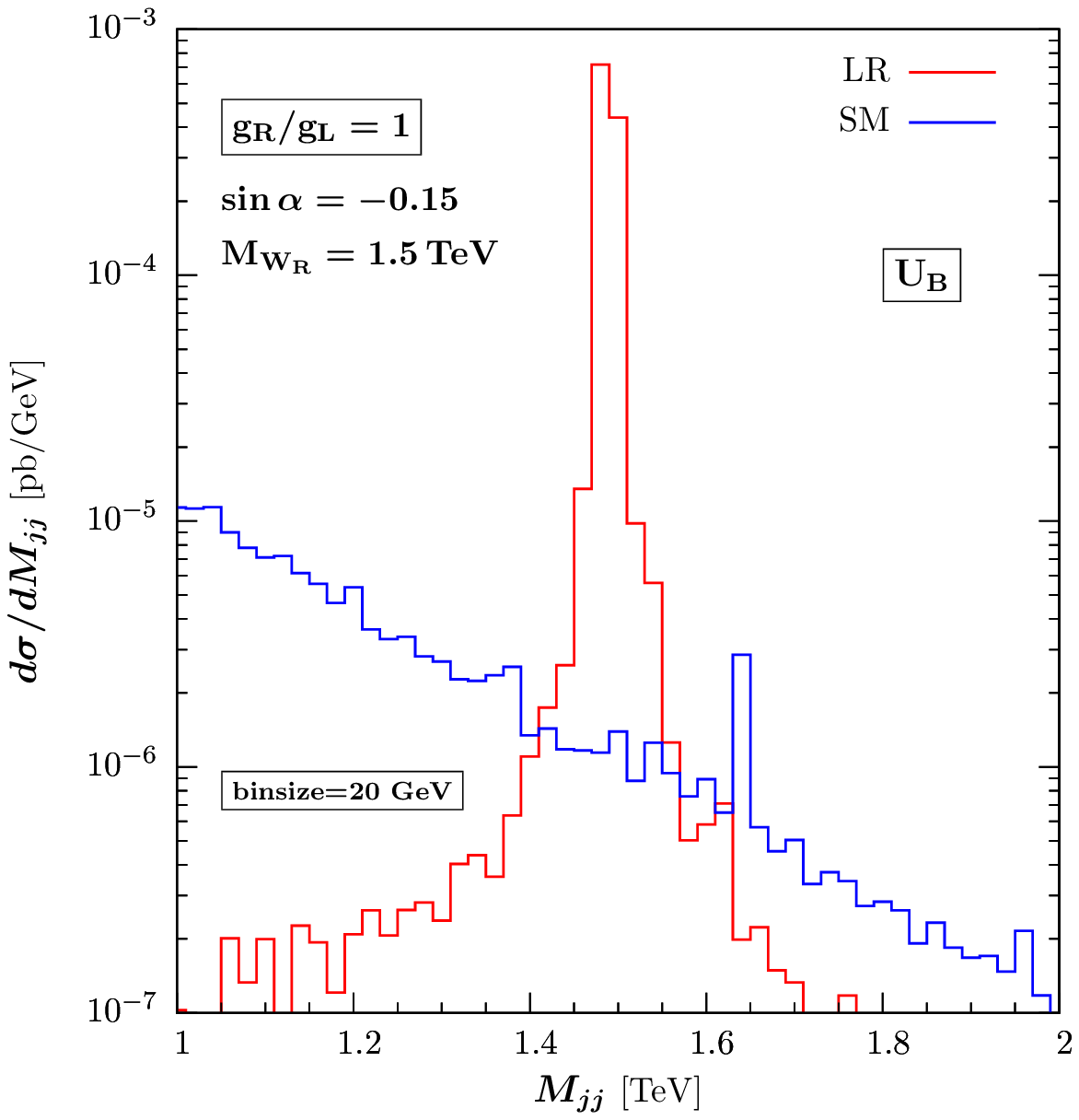} \\
 \includegraphics[scale=0.57]{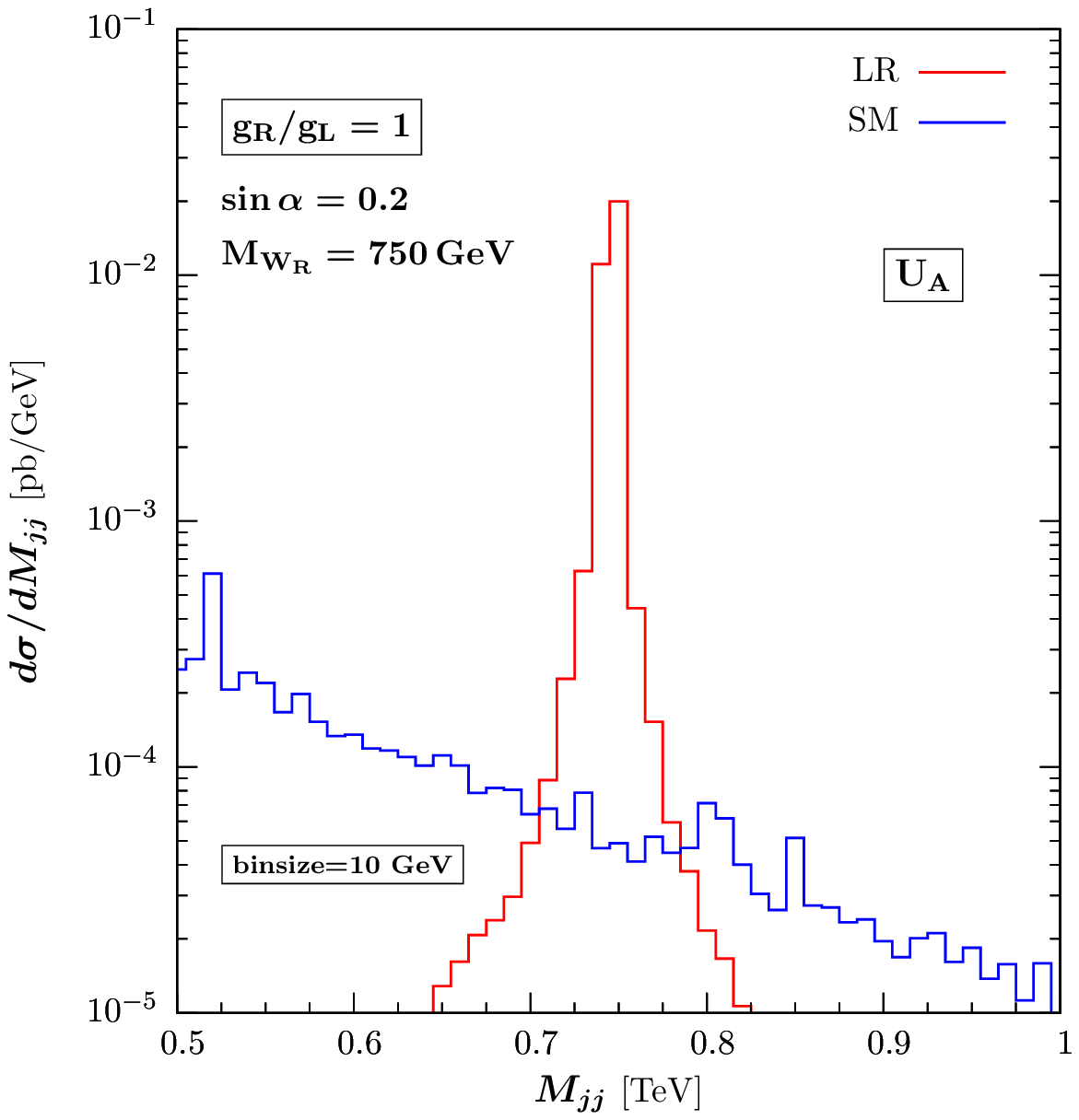} &
 \includegraphics[scale=0.57]{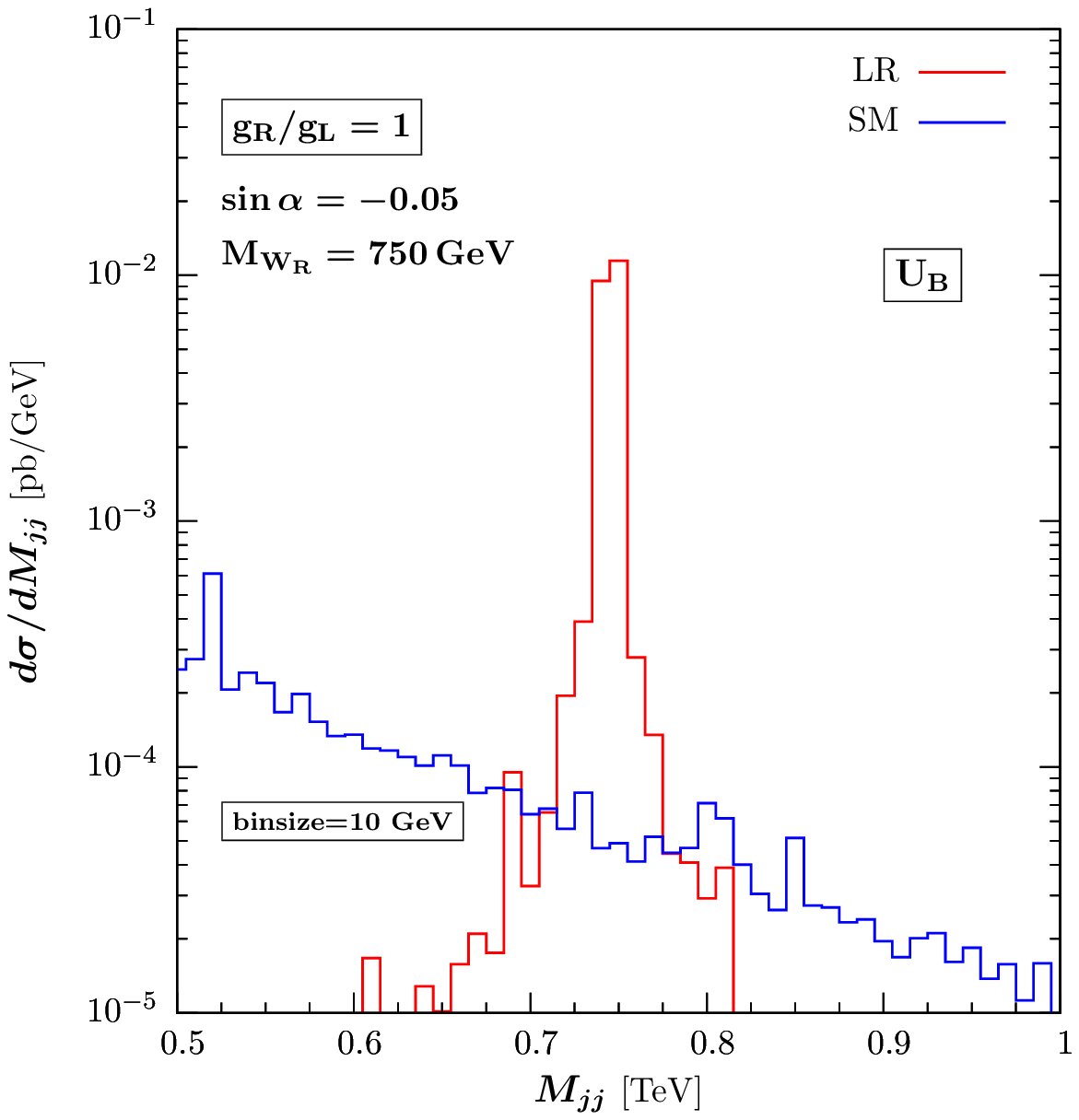}
\end{array}$
\caption{$W_R$ signal as a resonance in dijet mass distribution at the LHC
 with $U_A$ (left column) and $U_B$ (right column) right CKM parametrizations.
The signal is observed in $p,p \to t,dijet$ process where only the light quarks
are counted as jets. We choose $binsize=20$ GeV for intermediate $M_{W_R}=1500$
GeV and $binsize=10$ GeV for light $M_{W_R}=750$ GeV. The center of mass
energy is 14 TeV.}
\end{figure}

In Figure 7 and Figure 8, we present $W_R$ production signal at 14 TeV with
different
 CKM parametrizations and compare it  with the SM background. We choose the
binsize to be $20$ GeV, and plot the differential cross section with respect to
the invariant $dijet$ mass $M_{jj}$. It is clear that for all the
parametrizations, the $W_R$ signal can be observed as a resonance in the $dijet$
invariant mass distribution at the LHC and is quite distinguishable from the SM
background. The diagrams for the SM background are very similar to the ones in
Fig. 6, $W_R$ replaced with $W_L$ as well as some other exchange
diagrams. Signatures in $U_A$ and $U_B$ parametrizations are in the left and
right columns of Figure 7. In first two rows we kept $W_R$ mass at an
intermediate value ($M_{W_R}=1.5$ TeV) and changed the ratio of gauge coupling
constants ($g_R/g_L$) as well as the right CKM matrix element ($\sin\alpha$)
between the panels. The numerical values of these parameters are chosen
according to the constraints from low energy phenomenology in Figure 3. In the
last row we showed the signal of a lighter $W_R$ ($M_{W_R}=750$ GeV) with
binsize= 10 GeV and equal gauge coupling constants ($g_R=g_L$) in the region
allowed by the constraints. By comparison, in Figure 8 we show signatures for
the $W_R$ production and decay to $dijets$ in the MLRSM model for the
intermediate (left panel) and the light (right panel) $W_R$ (the last for
comparison only, as light $W_R$ masses are largely excluded by Kaon
phenomenology in the absence of extreme fine tuning). It is inferred from these
figures that a new right handed charged gauge boson signal of left-right
symmetry is very clear, distinct and accessible within the LHC's discovery
limits. For a luminosity of 100 fb$^{-1}$ at 14 TeV and a light $W_R$ boson
(both very optimistic assumptions), the signal can reach 100 events per year.

\begin{figure}[t]
\vspace{-0.4in}
\hspace{-0.5in}
$\begin{array}{cc}
 \includegraphics[scale=0.57]{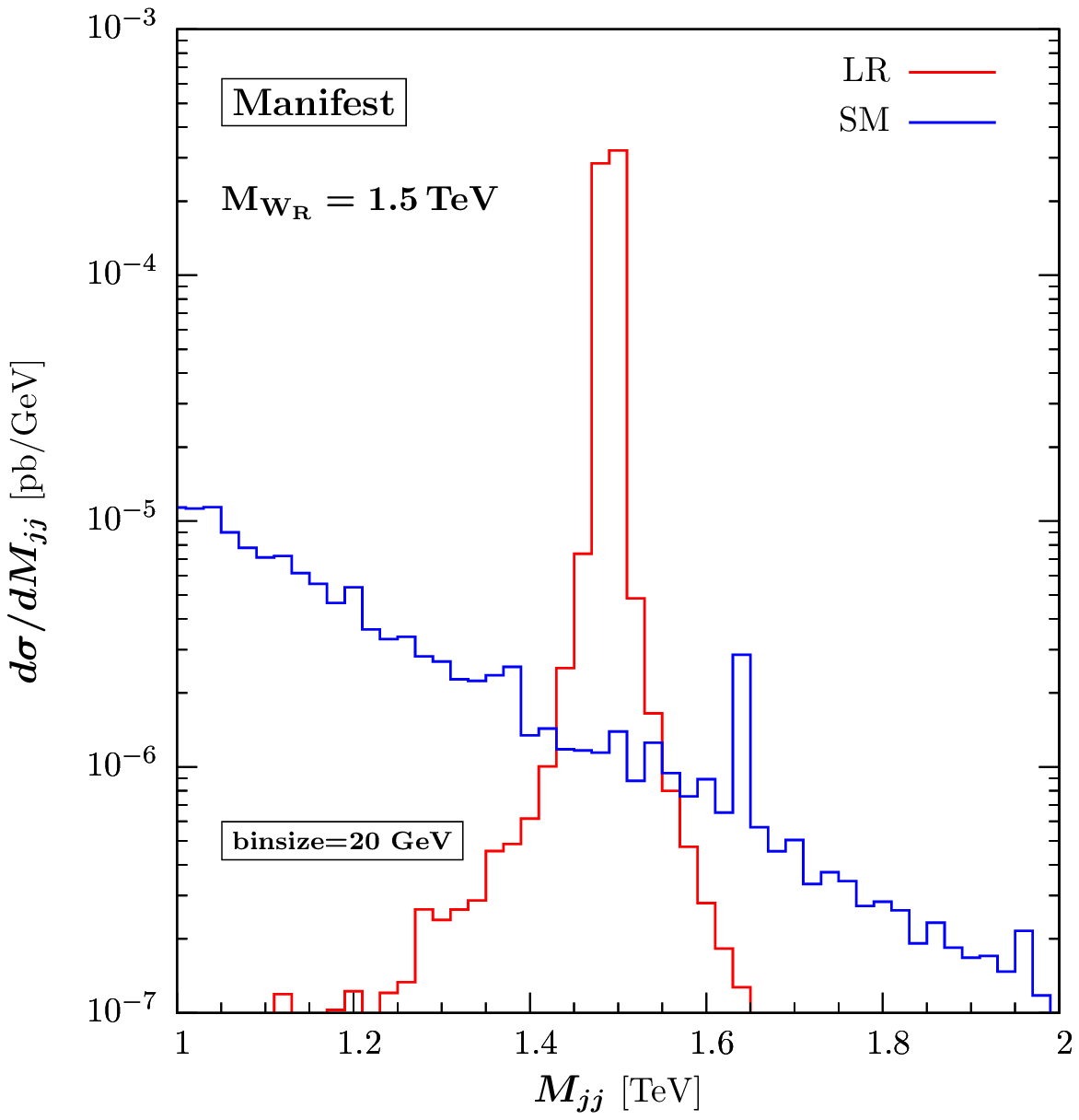} & 
 \includegraphics[scale=0.57]{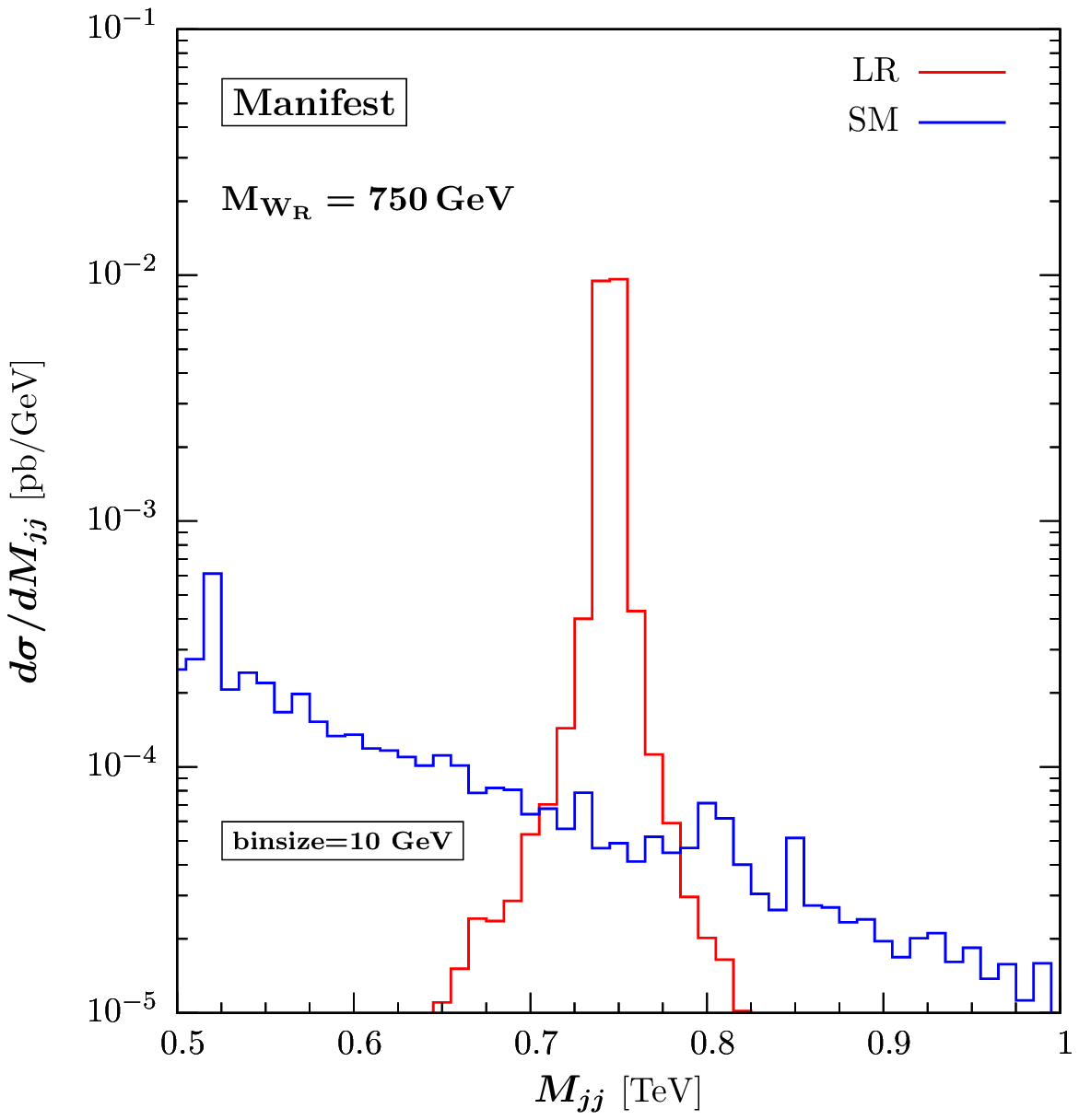}
\end{array}$
\caption{The resonance $W_R$ signal in the LHC with Manifest model.
 The intermediate $W_R$ (on left panel) and light $W_R$ (on right panel) signals
are presented. Again the same $binsize$ choice with Fig.7. The center of mass
energy is 14 TeV.}
\end{figure}

Whatever the model is, the cross-sections are robust, that is they are roughly
 the same order of magnitude, independent of the model used. The reason is the
following : in $U_A$ and $U_B$ models there are fewer diagrams contributing to
the differential cross sections, but the flavor violation from the right-handed
quarks is stronger, whereas in the manifest left-right symmetric model there are
more Feynman diagrams contributing to the differential cross section, but the
flavor violating interactions in the right-handed sector are weaker. This
explains the resemblance of the signals between $U_A(U_B)$ and the MLRSM. To
distinguish among the left-right models and to finely pinpoint the origin of the
signal requires further detailed analysis  with more realistic detector
simulations.

\section{Conclusions}
\label {conclude}

We analyzed the single production, decay and collider signals of $W_R$
 bosons produced in left-right symmetric models. We considered models with a
general right-handed quark mixing structure (which we call the asymmetric
left-right model), but constrained by the Kaon and B-meson flavor physics. We
also compared the results with those of the manifest left-right symmetric model,
where $V^R_{CKM}=V^L_{CKM}$ and the coupling constants in the left and right
sectors are equal.  In the asymmetric left right model, there is only one free
parameter in the right-handed CKM mixing matrix. Additionally, the charged Higgs
and $W_R$ masses, as well as the ratio of the $SU(2)_L$ and $SU(2)_R$ coupling
constants, are also free parameters. We included restrictions on the same
parameter space coming from $B_{d,s}^0-{\bar B}_{d,s}^0$ mixing and the
branching ratio for $b \to s \gamma$ \cite{Frank:2010qv}.

The dominant production mode is in association with a top quark, and this
 has a large background from the single top production (in association with
$W_L$) from the standard model. However looking at events in the 500-2000  GeV
mass range for $W_R$, we show that the SM background is always below the $W_R$
background, and we expect a significant peak above the SM background around the
$W_R$ mass (assumed to be in the range considered). Even with a luminosity of
$10$ fb$^{-1}$, achievable at the LHC within the next 3 years, we expect several
events a year, while with ${\cal L}=100$ fb$^{-1}$, the events could reach 100
per year. We concentrate our analysis in the $W_R \to dijet$ decay mode, where
$dijets$ are the light quarks $u,d,s$ and $c$. 

The cross section for the single $W_R$ production can reach $10$ fb,
 including all parameter restrictions, and the dominant decay modes are to light
quarks, ${\bar u}d(s)$ being favored by the choice of parametrization, and
${\bar c}s(d)$ and ${\bar t}b$ by the restrictions on the right handed CKM.

Models which predict extra $W^\prime$ bosons all have features that distinguish
 them from $W_R$ bosons in LRSM.   In warped extra-dimensional models \cite{RS},
the coupling of the extra charged gauge bosons to light quarks and leptons is
suppressed relative to those in SM. By contrast, in LRM, the decays to leptons
might be suppressed for heavy right-handed neutrinos, whereas  $W \to jet~jet$
has no missing energy so the signal can be reconstructed in full.  The
irreducible SM background from the electroweak process (single top production)
is shown to be smaller than the signal inside the resonance region. Warped RS
models need luminosities of ${\cal L}=100~ (1000)$  fb$^{-1}$ for a $W^\prime$
to reach a statistically significant signal, and expected $W^\prime$ masses are
in the 2-3 TeV region. 
  Technicolor or composite Higgs \cite{Higgless} models are expected to give
very
 similar signals, as the warped extra dimensional model is dual to the 4D strong
dynamics involved in electroweak symmetry breaking.
 In the Little Higgs Models \cite{Little Higgs}, the heavy $W_H$ is left-handed
 and the partial width to each fermion species is almost the same (for massless
fermions). In UED, the additional (KK) W and Z bosons expected to have masses in
the 100-200 GeV region \cite{UED}, have their hadronic decays closed, so they
decay democratically to all  lepton (one KK and one ordinary) flavors.
 
 A clear signal for a charged vector boson will be much more significant
 that one for a neutral $Z^\prime$ boson, as it would restrict the extension of
the gauge sector. Our analysis is complementary to previous analyses which
indicate how to find whether the extra charged $W^\prime$ boson is left or
right-handed, by presenting the signals expected for $W_R$ in LRM, both manifest
(with $V^R_{CKM}=V^L_{CKM}$) and in a case where $V^R_{CKM}$, constrained by B
and K phenomenology,  is independent on the mixing in the left-handed quark
sector and characterized by a single parameter. The signal for such a charged
gauge boson is significantly different than in other scenarios with extra
$W^\prime$s and would be an irrefutable signal of left-right symmetry.



\begin{thebibliography}{99}

\bibitem{Fukuda:1998mi}
  Y.~Fukuda {\it et al.}  [Super-Kamiokande Collaboration],
  Phys.\ Rev.\ Lett.\  {\bf 81}, 1562 (1998).

\bibitem{Pati:1974yy}
  J.~C.~Pati and A.~Salam,
  Phys.\ Rev.\  D {\bf 10}, 275 (1974)
  [Erratum-ibid.\  D {\bf 11}, 703 (1975)]; 
  R.~N.~Mohapatra and J.~C.~Pati,
  Phys.\ Rev.\  D {\bf 11}, 566 (1975); 
  R.~N.~Mohapatra and J.~C.~Pati,
  Phys.\ Rev.\  D {\bf 11}, 2558 (1975); 
  G.~Senjanovic and R.~N.~Mohapatra,
  Phys.\ Rev.\  D {\bf 12}, 1502 (1975); 
  R.~N.~Mohapatra, F.~E.~Paige and D.~P.~Sidhu,
   ``Symmetry Breaking And Naturalness Of Parity Conservation In Weak Neutral
  Phys.\ Rev.\  D {\bf 17}, 2462 (1978);  
  G.~Senjanovic,
  Nucl.\ Phys.\  B {\bf 153}, 334 (1979).


\bibitem{Mohapatra:1979ia}
  R.~N.~Mohapatra and G.~Senjanovic,
  Phys.\ Rev.\ Lett.\  {\bf 44}, 912 (1980).

  
 
\bibitem{Gunion:1989in}
  J.~F.~Gunion, J.~Grifols, A.~Mendez, B.~Kayser and F.~I.~Olness,
  Phys.\ Rev.\  D {\bf 40}, 1546 (1989); 
  J.~A.~Grifols,
  Phys.\ Rev.\  D {\bf 18}, 2704 (1978);  
  J.~A.~Grifols, A.~Mendez and G.~A.~Schuler,
  Mod.\ Phys.\ Lett.\  A {\bf 4}, 1485 (1989);  
  N.~G.~Deshpande, J.~F.~Gunion, B.~Kayser and F.~I.~Olness,
  Phys.\ Rev.\  D {\bf 44}, 837 (1991); 
  M.~L.~Swartz,
  Phys.\ Rev.\  D {\bf 40}, 1521 (1989); 
  R.~Vega and D.~A.~Dicus,
  Nucl.\ Phys.\  B {\bf 329}, 533 (1990); 
  K.~Huitu, J.~Maalampi, A.~Pietila and M.~Raidal,
  Nucl.\ Phys.\  B {\bf 487}, 27 (1997); 
  A.~Datta and A.~Raychaudhuri,
  Phys.\ Rev.\  D {\bf 62}, 055002 (2000); 
  G.~Barenboim, M.~Gorbahn, U.~Nierste and M.~Raidal,
  Phys.\ Rev.\  D {\bf 65}, 095003 (2002); 
  K.~Kiers, M.~Assis and A.~A.~Petrov,
  Phys.\ Rev.\  D {\bf 71}, 115015 (2005); 
  A.~G.~Akeroyd and M.~Aoki,
  Phys.\ Rev.\  D {\bf 72}, 035011 (2005).

  
 
\bibitem{Abazov:2004au}
  V.~M.~Abazov {\it et al.}  [D0 Collaboration],
  Phys.\ Rev.\ Lett.\  {\bf 93}, 141801 (2004); 
  D.~E.~Acosta {\it et al.}  [CDF Collaboration],
  Phys.\ Rev.\ Lett.\  {\bf 93}, 221802 (2004); 
  D.~E.~Acosta {\it et al.}  [CDF Collaboration],
  Phys.\ Rev.\ Lett.\  {\bf 95}, 071801 (2005).
  
  \bibitem{RS}
  K. Agashe, A. Delgado, M. J. May and R. Sundrum, JHEP 0308, 050 (2003); 
  K. Agashe et al., Phys.\ Rev.\  D {\bf 76}, 115015 (2007);
K. Agashe, S. Gopalakrishna, T. Han, G. Y. Huang and A. Soni, arXiv:0810.1497 [hep-ph].

\bibitem{UED}
 T. Appelquist, H. C. Cheng and B. A. Dobrescu, Phys.\ Rev.\ D {\bf 64}, 035002 (2001):
 H. C. Cheng, K. T. Matchev and M. Schmaltz, Phys.\ Rev.\ D {\bf 66},
056006 (2002). 


\bibitem{Little Higgs}
 N. Arkani-Hamed, A. G. Cohen, E. Katz and A. E. Nelson, JHEP 0207, 034 (2002); 
  D. E. Kaplan and M. Schmaltz, JHEP 0310, 039 (2003);
  T. Han, H. E. Logan, B. McElrath and L. T. Wang, Phys.\ Rev.\
D {\bf 67}, 095004 (2003). 

\bibitem{Higgless}
  C. Csaki, C. Grojean, L. Pilo and J. Terning, Phys. Rev. Lett. 92, 101802 (2004);
  R. S. Chivukula, B. Coleppa, S. Di Chiara, E. H. Simmons, H. J. He,
M. Kurachi and M. Tanabashi, Phys.\ Rev.\ D {\bf 74}, 075011 (2006):
H. J. He et al., Phys.\ Rev.\ D {\bf 78}, 031701 (2008).

\bibitem{Rizzo:1993fe}
  T.~G.~Rizzo,
  Phys.\ Rev.\  D {\bf 50} (1994) 325.
  
   
\bibitem{Gopalakrishna:2010xm}
  S.~Gopalakrishna, T.~Han, I.~Lewis, Z.~g.~Si and Y.~F.~Zhou,
  arXiv:1008.3508 [hep-ph];
  T.~G.~Rizzo,
  JHEP {\bf 0705}, 037 (2007).


\bibitem{Langacker:1989xa}
  P.~Langacker and S.~Uma Sankar,
  Phys.\ Rev.\  D {\bf 40}, 1569 (1989).

\bibitem{Amsler:2008zz}
  C.~Amsler {\it et al.}  [Particle Data Group],
  Phys.\ Lett.\  B {\bf 667} (2008) 1.

\bibitem{Tait:2000sh}
  T.~M.~P.~Tait and C.~P.~P.~Yuan,
  Phys.\ Rev.\  D {\bf 63}, 014018 (2001); 
  Q.~H.~Cao, J.~Wudka and C.~P.~Yuan,
  Phys.\ Lett.\  B {\bf 658}, 50 (2007).
  
   
  \bibitem{Tevatron} 
   V. M. Abazov et al. [D0 Collaboration],  Phys.\ Rev.\ Lett. {\bf 100}, 031804 (2008);  
 T. Aaltonen et al. [The CDF Collaboration], CDF Note 9246 (2008); 
 V. M. Abazov et al. [D0 Collaboration],  Phys.\ Rev.\ Lett. {\bf 100}, 211803 (2008). 
  
\bibitem{Frank:2010qv}
  M.~Frank, A.~Hayreter and I.~Turan,
  Phys.\ Rev.\  D {\bf 82}, 033012 (2010)
  [arXiv:1005.3074 [hep-ph]].

   
\bibitem{Senjanovic:1978ev}
  G.~Senjanovic,
  Nucl.\ Phys.\  B {\bf 153}, 334 (1979);  
  M.~A.~B.~Beg, R.~V.~Budny, R.~N.~Mohapatra and A.~Sirlin,
  Phys.\ Rev.\ Lett.\  {\bf 38}, 1252 (1977)
  [Erratum-ibid.\  {\bf 39}, 54 (1977)].
  
\bibitem{Harari:1983gq}
  H.~Harari and M.~Leurer,
  Nucl.\ Phys.\  B {\bf 233}, 221 (1984).
  
\bibitem{Kiers:2002cz}
  K.~Kiers, J.~Kolb, J.~Lee, A.~Soni and G.~H.~Wu,
  Phys.\ Rev.\  D {\bf 66}, 095002 (2002)
  [arXiv:hep-ph/0205082].

  
\bibitem{Langacker:1984dp}
  P.~Langacker,
  Phys.\ Rev.\  D {\bf 30}, 2008 (1984).
  
\bibitem{Ma:1992ve}
  W.~Ma, X.~Li and Y.~Liu,
  Phys.\ Rev.\  D {\bf 45} (1992) 1792; 
  J.~Maalampi, A.~Pietila and J.~Vuori,
  Nucl.\ Phys.\  B {\bf 381}, 544 (1992); 
  P.~Langacker, R.~W.~Robinett and J.~L.~Rosner,
  Phys.\ Rev.\  D {\bf 30}, 1470 (1984).

\bibitem{Barenboim:1996nd}
  G.~Barenboim, J.~Bernabeu, J.~Prades and M.~Raidal,
  Phys.\ Rev.\  D {\bf 55}, 4213 (1997).

  
  
\bibitem{Hou:1985ur}
  W.~S.~Hou and A.~Soni,
  Phys.\ Rev.\  D {\bf 32}, 163 (1985); 
  L.~Maharana,
  Phys.\ Lett.\  B {\bf 149}, 399 (1984); 
  J.~M.~Frere, J.~Galand, A.~Le Yaouanc, L.~Oliver, O.~Pene and J.~C.~Raynal,
  Phys.\ Rev.\  D {\bf 46}, 337 (1992); 
  D.~Chang, J.~Basecq, L.~F.~Li and P.~B.~Pal,
  Phys.\ Rev.\  D {\bf 30}, 1601 (1984); 
  G.~Beall, M.~Bander and A.~Soni,
  Phys.\ Rev.\ Lett.\  {\bf 48}, 848 (1982); 
  R.~N.~Mohapatra, G.~Senjanovic and M.~D.~Tran,
  Phys.\ Rev.\  D {\bf 28}, 546 (1983);  
  P.~Colangelo and G.~Nardulli,
  Phys.\ Lett.\  B {\bf 253}, 154 (1991).
 
\bibitem{Maiezza:2010ic}
  A.~Maiezza, M.~Nemevsek, F.~Nesti and G.~Senjanovic,
  arXiv:1005.5160 [hep-ph].
 
  
\bibitem{Gilman:1983ce}
  F.~J.~Gilman and M.~H.~Reno,
   ``Restrictions From The Neutral K And B Meson Systems On Left-Right Symmetric
  Phys.\ Rev.\  D {\bf 29}, 937 (1984); 
  L.~Maharana, A.~Nath and A.~R.~Panda,
  Phys.\ Rev.\  D {\bf 47} (1993) 4998.

\bibitem{Silverman:1997fx}
  D.~Silverman and H.~Yao,
  JHEP {\bf 0110}, 008 (2001).
  
\bibitem{Rizzo:1994aj}
  T.~G.~Rizzo,
  Phys.\ Rev.\  D {\bf 50}, 3303 (1994); 
  G.~Bhattacharyya and A.~Raychaudhuri,
  Phys.\ Lett.\  B {\bf 357}, 119 (1995); 
  C.~S.~Kim and Y.~G.~Kim,
  Phys.\ Rev.\  D {\bf 61}, 054008 (2000).


\bibitem{Beall:1981zq}
  G.~Beall and A.~Soni,
  Phys.\ Rev.\ Lett.\  {\bf 47}, 552 (1981); 
  J.~M.~Frere, J.~Galand, A.~Le Yaouanc, L.~Oliver, O.~Pene and J.~C.~Raynal,
  Phys.\ Rev.\  D {\bf 45}, 259 (1992); 
  M.~E.~Pospelov,
  Phys.\ Rev.\  D {\bf 56}, 259 (1997).

 \bibitem{CDFdijet}
 T.~Aaltonen  {\it et. al} [CDF Collaboration], Phys.\ Rev.\  D {\bf 79}, 112002 (2009).
  
\bibitem{Pukhov:2004ca}
  A.~Pukhov,
  arXiv:hep-ph/0412191.
    

\end{thebibliography}
\end{document}